\documentclass[pra,twocolumn,amsmath,amssymb,floatfix]{revtex4}

\usepackage{amsmath}
\usepackage{graphicx}
\usepackage{epsfig}
\usepackage{hyperref}
\hypersetup{}
\usepackage{braket}
\usepackage{tikz}

\usepackage{bm}

\def\H1{\widehat{H}_1}

\newcommand{\be}{\begin{equation}}
\newcommand{\ee}{\end{equation}}
\newcommand{\beq}{\begin{eqnarray}}
\newcommand{\eeq}{\end{eqnarray}}

\begin{document}

\title[]{Photonic Kondo model}

\author{Mikhail Pletyukhov$^{1}$, Niclas M\"{u}ller$^1$, Vladimir Gritsev$^{2}$ }

\affiliation{
$^1$Institute for Theory of Statistical Physics and JARA -- Fundamentals of Future Information Technology, 
RWTH Aachen University, 52056 Aachen, Germany \\
$^2$Institute for Theoretical Physics, Universiteit van Amsterdam, Science Park 904,
Postbus 94485, 1098 XH Amsterdam, The Netherlands
}
\begin{abstract}
Here we introduce and study a photonic analogue of the Kondo model. The model is defined as a far detuned regime of photonic scattering off a three-level emitter in a $\Lambda$-type configuration coupled to a one-dimensional transmission line with linear dispersion. We characterize this system by studying the dynamics of a local system (emitter) driven by a coherent field pulse as well as the first- and second-order correlation functions of the scattered light. Various polarization-dependent correlation effects including entanglement between the emitter and transmitted photons are quantified by the purity of the emitter's state. We also show that statistical properties of the scattered light are very sensitive to the polarization of the incoming coherent pulse. 
\end{abstract}

\maketitle

\section{Introduction}

In the last years there were many theoretical proposals and their partial experimental realizations which are aimed to bride the interface between condensed matter physics and quantum optics. The objective of these proposals is twofold. Some of them are triggered by ideas of quantum simulation of quantum many-body condensed matter models using the tools of quantum optics while the others were inspired by advances in quantum information, since the eventual scalability of quantum architecture would inevitably introduce some many-body aspects into quantum optical dynamics. 

On the quantum simulation side these proposals include, in particular,  the Bose-Hubbard model of Mott-superfluid transition for polaritons \cite{BH1},\cite{BH2}, one-dimensional physics leading to photonic fermionization \cite{LL1}, \cite{LL2}, Bose condensation of light \cite{BEC}, topological photonic states \cite{top1},\cite{top2} and quantum Hall fluids for photons \cite{Hafezi1}. For a recent overview see \cite{CiutiRMP}.

On the quantum information and computation side the main player is entanglement which requires the presence of interaction between elements of quantum networks with simultaneous protection against the action of the environment \cite{Kimble}. In this respect several proposals of hybrid systems are considered as the most promising route for realizing future functional devices. The role of qubits in these schemes are played by real or artificial atoms (e.g. quantum dots) or other solid-state based structures (e.g. NV-centers, for a review see \cite{ChuLukin}) while the information between them is transferred by photons or other (sometimes collective) excitations (e.g. polaritons). To ensure efficient functionality, qubits must be entangled in a controllable way. Entanglement can be encoded into either polarization degrees of freedom of photons or some collective degrees of freedom,  and can be shared with internal  degrees of freedom of the qubit (e.g. spin), thus producing qubit-field entanglement which may then be transferred to a different qubit of a device by photons. This philosophy has been successfully realized in several recent experiments thus achieving qubit-qubit entanglement for distances ranging from nanometers to kilometers \cite{Neumann}, \cite{Ansmann}, \cite{JW-room}, \cite{Bernien}.

\begin{figure}[ht]
\centering
\includegraphics[width=86mm]{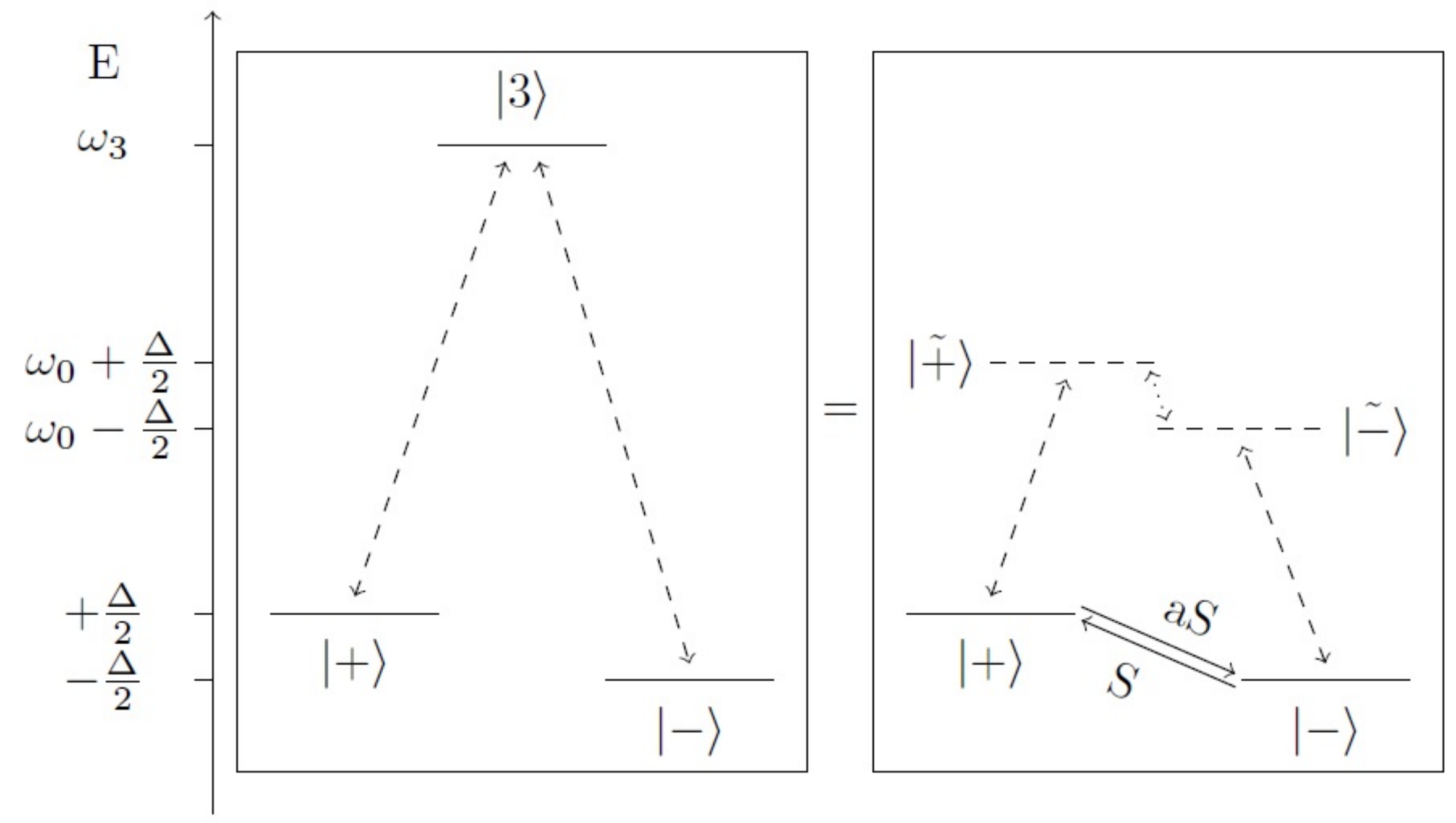}
    \caption{{\it Left panel:} Level structure of the local far-detuned $\Lambda$-system. The dashed arrows represent  transitions allowed by means of single-photon processes. Direct transitions between the states $|+\rangle$ and $|-\rangle$ are forbidden by angular momentum selection rules. {\it Right panel:}  An equivalent description in terms of two-photon scattering via the virtual states $|\tilde{\pm}\rangle$ gives rise to an effective two level model with Stokes (S) and anti-Stokes (aS) processes, which are represented by the solid arrows. Here $\omega_0$ denotes the energy of incident radiation, and $\Delta$ denotes the two-photon detuning. } 
 \label{scheme}
\end{figure}

Here we present a model which, on one hand, extends the list of proposals for photonic quantum simulators, and on the other hand can be used for quantum information purposes as an interface for spin-qubit entanglement transfer. The effect we are discussing here bears a number of similarities with a Kondo effect known in condensed matter physics, while many important differences remains because of Bose statistics of photons. With these differences in mind we call it {\it photonic Kondo model}, in analogy with the fermionic Kondo model. The electronic Kondo model, as introduced by J. Kondo in 1964, \cite{Kondo}, describes the interaction of a reservoir of electrons coupled to a quantum impurity modeled by a local spin-$\frac{1}{2}$ system. It was proposed to account for a nonmonotonous behavior of the resistivity in certain metallic alloys containing small concentrations of magnetic impurities at low temperatures. The model can, in a certain limit, be derived from the Anderson model \cite{Anderson}. Later on, it has been extended to problems of nonequilibrium electronic transport through quantum dots \cite{Cronenwett}, \cite{Glazman}. From the quantum optical point of view the ideology of impurity models is quite natural, since the localized quantum optical emitters resemble local impurity spins of condensed matter models while propagating photons are the analogues of itinerant electrons in solids. 

We note that the model we introduce and study here should be distinguished from the other quantum optical models studied in the literature which are somewhat related to the Kondo model. First, the strong coupling regime of the one-dimensional transmission line coupled to the two level system has been mapped to the {\it bosonized} form of the anisotropic Kondo model \cite{Le1}, \cite{KoLe}, \cite{LLLS}, \cite{Le2}. This mapping has been extended and studied in relation with a microwave realization of the ohmic regime in the spin-boson model in \cite{LeHur1}, \cite{LeHur2}. Finally, our model is completely different from the photon-assisted Kondo effect studied in \cite{Tureci1}, \cite{Tureci2}.   

One of the possible realizations of our model can be obtained by using a three-level $\Lambda$ scheme (see Fig.~\ref{scheme}) where single-photon transitions $|+\rangle\leftrightarrow |3\rangle$ and  $|-\rangle\leftrightarrow |3\rangle$ are only coupled to photons of left  ($\sigma =-$) and right ($\sigma=+$) polarizations, respectively. The $|+\rangle\leftrightarrow |-\rangle$ transition is optically forbidden by the angular momentum selection rules. The qubit-photon entanglement is achieved by de-exciting the qubit prepared initially in the state $|3\rangle$, thus creating a quantum superposition of states $|+\rangle|-\rangle_{ph}$ and $|-\rangle|+\rangle_{ph}$.

From a more general perspective, multi-level schemes exhibit a variety of linear and nonlinear properties built in by quantum interference phenomena between different quantum level pathways. This includes coherence population trapping, electromagnetically induced transparency \cite{LukinEIT}, \cite{FIM}, coherent population trapping \cite{AGMO}, \cite{AO}, \cite{GWS}, \cite{Orriols}, \cite{Arimondo}, \cite{Aspect}, stimulated rapid adiabatic passage (STIRAP) \cite{stirap1}, \cite{stirap2}, \cite{stirap-rev1}, \cite{stirap-rev2}, \cite{stirap-rev4}, Autler-Tawnes effect \cite{AT}, \cite{AT-CT}, \cite{EIT-vs-AT}, resonance fluorescence \cite{CT-R}, \cite{WS}, \cite{Sobol}, \cite{Fu}, \cite{Mavroyannis1}, \cite{Mavroyannis2}, \cite{AB}, \cite{BPK}, \cite{exp-qd}, controllable Kerr nonlinearity \cite{HH}, \cite{Kash},\cite{WGX}, two-photon fluorescence \cite{DSW}, \cite{2-photon}. In addition, the role of photonic polarization essentially increases in the Raman and magneto-optical effects \cite{Budker}. 

The photonic Kondo model which we introduce in this paper is an effective model to describe dynamics of the local  three-level system in a regime where the two-photon processes dominate. It allows us to solve exactly the dynamics of the scattered field and the local system, and to compute the first- and the second order correlation functions, including the polarization resolved inelastic power spectrum. In terms of this model we show that in the stationary limits these field observables can be related to the purity of the local system.

\section{Model} 
\subsection{General considerations\label{Consider}}
Let us start our consideration from the three level $\Lambda$-scheme in  a $\Lambda$ configuration shown in Fig. \ref{scheme}, left panel. It is irradiated by a coherent light of an arbitrary elliptic polarization, i.e. a superposition of left and right polarized photons, and of a frequency $\omega_0$ which is half as big as the frequency of the single-photon transitions $|+\rangle\leftrightarrow |3\rangle$ and  $|-\rangle\leftrightarrow |3\rangle$.

While in this far detuned regime single-photon processes are blocked by the energy conservation, the dominant processes will involve two photons which are scattered via the virtual states $| \tilde{\pm} \rangle$ (transitions to these states are characterized by the same quantum numbers as to the state $| 3\rangle$, besides the transition frequencies). In total, there are  four possible second-order virtual processes: (i) $|+\rangle \rightarrow |\tilde{+}\rangle \rightarrow |\tilde{-}\rangle\rightarrow |-\rangle$; (ii)
$|-\rangle \rightarrow |\tilde{-}\rangle \rightarrow |\tilde{+}\rangle\rightarrow |+\rangle$; (iii) $|+\rangle \rightarrow |\tilde{+}\rangle \rightarrow |+\rangle$; and (iv) $|-\rangle \rightarrow |\tilde{-}\rangle \rightarrow |-\rangle$. Here the "subprocesses" $|\tilde{\pm}\rangle \rightarrow |\tilde{\mp}\rangle$ symbolically mean an infinitesimally short-timed occupation of the energetically inaccessible state $| 3 \rangle$.

In the processes (i)  and (ii) one can recognize the inelastic Raman scattering, where the scattered photon changes its energy: (i) either it gains energy $\Delta$ (anti-Stokes scattering), or (ii) and it looses energy $\Delta$ (Stokes scattering). The parameter $\Delta$ is called the two-photon detuning. Both processes are accompanied by the change of the incoming photon polarization. 

The processes (iii) and (iv) occur without energy and polarization change, and correspond to the Rayleigh scattering. 

From this consideration we conclude that the power spectrum of this model exhibits three peaks: one central peak at $\omega = \omega_0$ and two-side peaks at $\omega = \omega_0 \pm \Delta$. This qualitative analysis, however, misses lineshapes of these transitions, as well as the precise value (denoted by $\Omega$ in the following) of the Stokes and anti-Stokes shifts, which differs from $\Delta$ by the field renormalization effects. In addition to the elastic Rayleigh peak, as we will show in the following, there is an inelastic, i.e. broadened, peak at $\omega = \omega_0$.

To quantitatively describe all these effects we need an effective model describing the transitions between the states $|+ \rangle $ and $| - \rangle$, that is operating in terms of a two-level system. Its derivation is the subject of the following subsection.

We note that interesting phenomena can arise only at $\Delta \neq 0$: for $\Delta =0$ inelastic processes cannot take place, and therefore no entanglement between the local system and the scattered field is to be expected.

\subsection{Photonic Kondo model: derivation}

The Hamiltonian describing the three level system sketched in Fig. \ref{scheme}, left panel,  reads
\begin{align}
H^{(3)} &= \omega_3 |3\rangle\langle 3| + \Delta S_z + \sum_{\sigma=\pm} \int d \omega \, \omega a^{\dagger}_{\omega \sigma} a_{\omega \sigma}
\label{dr1} \\
&+ \sum_{\sigma=\pm} \int d \omega g_{\sigma} [\sigma a_{\omega \sigma} \ket{3} \bra{-\sigma} + h.c. \rbrack,
\label{dr2} 
\end{align}
where $S_z \equiv \frac{1}{2} (\ket{+}\bra{+} - \ket{-}\bra{-})$, and $a^{\dagger}_{\omega \sigma}$ and $a_{\omega \sigma}$ are canonical bosonic creation and annihilation operators which create/destroy a photon with frequency $\omega$ and either right polarization ($\sigma=+$) or left polarization ($\sigma=-$). The first line, \eqref{dr1}), contains the free Hamiltonians of the local three-level system and of the waveguide radiation modes, while the second line, Eq. \eqref{dr2} contains the dipole-interaction term in the rotating wave approximation. The latter induces transitions $\ket{-}\leftrightarrow \ket{3}$ and $\ket{+} \leftrightarrow \ket{3}$ 
of the local system by means of absorption/emission of photons with right and left polarization, respectively. Matrix elements connecting the states $\ket{-}$ and $\ket{+}$ vanish identically in accordance with the angular momentum selection rules, as  stated in the Introduction. 

When a driving field is far-off-resonance to the transitions $\ket{-}\leftrightarrow \ket{3}$ and $\ket{+} \leftrightarrow \ket{3}$, the state  
$\ket{3}$ becomes virtual, i.e. it can only be occupied during infinitesimally short time. Then, the energy conserving processes are of the second order in the coupling $g$, i.e. they involve two photons. 

To make this explicit we perform the (unitary) Schrieffer-Wolff transformation \cite{SchriefferWolff},
\begin{align}
H^{(3)} \mapsto e^S H^{(3)} e^{-S} &= H^{(3)} + \lbrack S , H^{(3)} \rbrack \nonumber\\
&+ \frac{1}{2} \lbrack S , \lbrack S , H^{(3)} \rbrack \rbrack + \ldots \, ,
\label{dr5}
\end{align}
which eliminates single-particle processes and gives rise to an effective -- cotunneling -- model. To achieve this, it is necessary to choose the generator $S$ in the form
\begin{align}
S &= \sum_{\sigma = \pm } \int d \omega \, \Lambda_{\omega \sigma} [\sigma a_{\omega \sigma} \ket{3} \bra{-\sigma} - h.c.]
\label{S}
\end{align}
with $\Lambda_{\omega \sigma} = \frac{g_{\sigma}}{\omega_3 - \omega} \approx \frac{2 g_{\sigma}}{\omega_3}$ (assuming $\omega \approx \frac{\omega_3}{2}$). Truncating (\ref{dr5}) after the double commutator and projecting the result onto the subspace of the local system spanned by the states $\ket{+}$ and $\ket{-}$, we obtain for $g_+ = g_- \equiv g$ (see Appendix \ref{App:AppendixA} for details of the derivation) an effective two-level Hamiltonian
\begin{align}
H^{(2)} &= \Delta S_z + \sum_{\sigma} \int d \omega \, \omega a^{\dagger}_{\omega \sigma} a_{\omega \sigma} \nonumber \\
&+ 2 \pi  J \left[ \boldsymbol{s}(0) \cdot \boldsymbol{S} - \frac{n(0)}{4} \right],
\label{dr8}
\end{align}
where 
\begin{align}
J &= \frac{4 g^2}{\omega_3} >0
\label{dr9} 
\end{align}
is the ``antiferromagnetic'' exchange coupling between the local ``spin-$\frac12$''
\begin{align}
S^i &= \sum_{\sigma, \sigma'}  \ket{\sigma} \frac{\sigma^i_{\sigma \sigma'}}{2} \bra{\sigma^{\prime}}
\label{dr9a}
\end{align}
and the ``reservoir spin density''
\begin{align}
s^i(0) &= \sum_{\sigma , \sigma'} \int \frac{d \omega d\omega^{\prime}}{2 \pi} a^{\dagger}_{\omega \sigma } \frac{\sigma^i_{\sigma \sigma^{\prime}} }{2} a_{\omega^{\prime}\sigma^{\prime}  }.
\label{dr10} 
\end{align}
Both $S^i$ and $s^i(0)$ are expressed in terms of the Pauli matrices $\sigma^i$, $i=x,y,z$. Additionally, the ``reservoir particle density'' is defined by
\begin{align}
n(0) &= \sum_{\sigma} \int \frac{d\omega d\omega^{\prime}}{2 \pi}a^{\dagger}_{\omega \sigma} a_{ \omega^{\prime} \sigma}.
\label{dr11}
\end{align}

Apart from the difference in statistics of particles, the Hamiltonian \eqref{dr8} represents the (isotropic) Kondo model, and therefore it can be named a bosonic Kondo Hamiltonian. It can be used for a description of the coherently driven $\Lambda$-atom in the far-detuned regime in terms of a pseudo-spin-$\frac{1}{2}$ system which is "antiferromagnetically" coupled to the polarization density of a bosonic bath. Note that for $g_+ \neq g_-$ one obtains the anisotropic Kondo model with $J_{\parallel} =\frac{2 (g_+^2 + g_-^2)}{\omega_3}$ and $J_{\perp}= \frac{4 g_+ g_-}{\omega_3}$, which however won't be studied in this paper.

Despite the similarity of the Hamiltonians, there are also essential differences between the electronic (fermionic) and the photonic (bosonic) Kondo models. 

First of all, the fields in the two models obey different kinds of statistics. In the fermionic case, the Pauli exclusion principle restricts the set of possible states, while there is no such restriction in the bosonic case. Hence, the initial states that are considered in the two models are significantly different. In the electronic Kondo model, as an initial state one usually considers the Fermi sea (at zero temperature), while in the photonic case one often chooses a single-mode coherent state with a large mean number of photons in order to describe a laser field. Due to the properties of a coherent state one can obtain an {\it exact analytic solution} of the Heisenberg equations of motion in the photonic Kondo model, which is hardly possible in its electronic counterpart.

Secondly,  observables and methods of their measurements are also different. In the electronic case one typically measures a current through the sample (for more than one attached reservoir) and its autocorrelations and noise spectra. In the photonic case one can measure both the average field and its correlation functions, which characterize the statistical properties of the electromagnetic field after interaction, as will be shown below.

For these reasons we do not associate any kind of the {\it Kondo effect} to the photonic Kondo model, in the sense how this effect is detected and interpreted in the electronic context (e.g., in a form of the quasiparticle -- Kondo -- resonance, being observed in the differential conductance in the zero bias limit and having a characteristic scale, called the Kondo temperature).

There are, however, some general features shared by the electronic and the photonic Kondo models: (i) one can still interpret virtual cotunneling processes in the same terms, as was discussed in Sec. \ref{Consider}; (ii) the two models basically describe the same kind of interaction, namely the exchange coupling of a local (pseudo-)spin-$\frac{1}{2}$  to a spin density  of bath states, implying that in both cases scattering occurs due to spin fluctuations.

\subsection{Time evolution of the local system} 
Using the effective two-level Hamiltonian \eqref{dr8} and the canonical commutation relations of the bosonic field operators, we find the Heisenberg equations of motion for the field operators
\begin{align}
\frac{d}{dt} a_{\omega \sigma} (t) = - i\omega a_{\omega \sigma}(t) - i J\sum_{\sigma'} \int d\omega^{\prime} \, M_{\sigma \sigma^{\prime}}(t) a_{\omega^{\prime} \sigma^{\prime} }(t),
\label{dr12}
\end{align}
where 
\begin{align}
M_{\sigma \sigma^{\prime}}(t) = \frac{\sigma^i_{\sigma \sigma^{\prime}}}{2} S^i (t) - \frac{ \delta_{\sigma \sigma^{\prime}}}{4}.
\label{dr13}
\end{align}

Assuming $t > x > 0$ we derive (see Appendix \ref{App:EqM} for details) the relation
\begin{align}
a_{\sigma}(x,t) &= \frac{1}{\sqrt{2 \pi}} \int d \omega a_{\omega \sigma} (t) e^{i \omega x} \label{dr14} \\
&= \sum_{\sigma'} [e^{i \phi} P^s_{\sigma \sigma^{\prime}}(t-x) + P^t_{\sigma \sigma^{\prime}}(t-x) ] a_{\sigma^{\prime}}(x-t,0),
\notag
\end{align}
where $a_{\sigma}(x,t)$ is the coordinate representation of the annihilation operator (the phase velocity is set to unity for convenience). The phase $\phi = 2 \arctan \pi J$ separates the  projectors
\begin{align}
P^s_{\sigma \sigma^{\prime}}(t) &= - M_{\sigma \sigma^{\prime}}(t), \\
P^t_{\sigma \sigma^{\prime}}(t) &= \delta_{\sigma \sigma^{\prime}} + M_{\sigma \sigma^{\prime}}(t),
\end{align}
onto the singlet ($P^s$) and the triplet ($P^t$) configurations, which are formed of the local spin and the itinerant spin density.

The result \eqref{dr14} is the central equation in our calculation, because it links dynamics of the scattered field to dynamics of the local system and to an initial state of the field. Specifying further the initial state of the field allows us to express field-field correlations through spin-spin correlations.

From  \eqref{dr8} and \eqref{dr14}  we deduce the Heisenberg equation of motion for the local spin operators 
\begin{align}
& \frac{d}{dt} S^i(t) = 2 \pi  J \epsilon_{i j k} \sum_{\sigma, \sigma'} a^{\dagger}_{\sigma}(0,t) \frac{\sigma^j_{\sigma \sigma^{\prime}}}{2} S^k(t) a_{\sigma^{\prime}}(0,t) \nonumber\\
&  \qquad + \Delta \epsilon_{i z k} S^k(t) 
\notag \\
&= 2 \pi  J \cos^2 \frac{\phi}{2} \sum_{\sigma, \sigma'} a^{\dagger}_{\sigma}(- t,0) \epsilon_{ijk} \frac{\sigma^j_{\sigma \sigma^{\prime}}}{2} S^k (t) a_{\sigma^{\prime}}(-t,0) \notag \\
&+  \frac{\pi  J \sin \phi}{2} \sum_{\sigma, \sigma'} a^{\dagger}_{\sigma}(-t,0) \left[ \frac{\sigma^i_{\sigma \sigma^{\prime}}}{2} - \delta_{\sigma \sigma^{\prime}} S^i (t) \right] a_{\sigma^{\prime}}(-t,0) \notag \\
&+ \Delta \epsilon_{izk}  S^k (t).
\label{dr15}
\end{align}

To analyze dynamics of the local system on the basis of Eq. \eqref{dr15},  we need to explicitly define the field's initial state. 

\subsection{Dynamics starting from the initial coherent state \label{coherence}}
We assume the field to be initially in a coherent state with frequency $\omega_0$ and an arbitrary elliptic polarization. 
It is generated by the wavepacket operator, e.g. of a rectangular shape,
\begin{align}
& D  | vac\rangle = \prod_{\sigma = \pm} D_{\sigma}  | vac\rangle \notag \\
&= \prod_{\sigma= \pm} \mathcal{N_{\sigma}} \exp \left[ \frac{\alpha_{\sigma}}{\sqrt{L}} \int^{L/2}_{-L/2} dx \, e^{i \omega_0 x } a^{\dagger}_{\sigma}(x) \right] | vac\rangle ,
\label{dr16} 
\end{align}
where $\mathcal{N}_+ \mathcal{N}_-$ is a normalization constant,
$L$ is a spatial length of the rectangular pulse, and $\alpha_{\pm}$ are coherence parameters of the corresponding polarization. In the frequency representation, the wavepacket operator generating  the state \eqref{dr16} is centered around the frequency $\omega_0$ and broadened by the width $\frac{2 \pi}{L}$.
 
In terms of the coherence parameters $\alpha_{\pm}$, a polarization of the initial coherent state  \eqref{dr16} is given by the classical Jones vector \cite{Jones}
\begin{align}
\boldsymbol{s}_{cl} = \sum_{\sigma, \sigma'}  \alpha^*_{\sigma} \frac{\boldsymbol{\sigma}_{\sigma \sigma^{\prime}}}{2L}\alpha_{\sigma^{\prime}} = \frac{f}{2} \boldsymbol{n}_{cl},
\label{dr21} 
\end{align}
where $\boldsymbol{n}_{cl}$ is a unit vector, and $f = \sum_{\sigma} \frac{|\alpha_{\sigma}|^2}{L}$ is the photon density. In particular, the Jones vector parallel to $\pm \boldsymbol{e}_z$ corresponds to right/left-circularly polarized light, while the Jones vectors in the $x-y$ plane correspond to linear polarizations. Any other direction corresponds to an elliptic polarization.

In averaging Eq. \eqref{dr15} over the initial state  \eqref{dr16} (taken in the narrow bandwidth  limit $L \to \infty$) we use the property
\beq
a_{\sigma}(x,0) D_{\sigma} | vac \rangle =\frac{\alpha_{\sigma}}{\sqrt{L}} e^{i \omega_0 x}  D_{\sigma } | vac \rangle ,
\eeq
and obtain the equation of motion for the spin expectation values
\begin{equation}
\frac{d}{dt} \braket{\boldsymbol{S}(t)} = \boldsymbol{h}_{eff} \times \braket{\boldsymbol{S}(t)} - \Gamma \left[ \braket{\boldsymbol{S}(t)} - \frac{\boldsymbol{n}_{cl}}{2} \right],
\label{dr19}
\end{equation}
where
\begin{align}
\boldsymbol{h}_{eff} &= \pi  J f  \cos^2 \frac{\phi}{2} \boldsymbol{n}_{cl} + \Delta \boldsymbol{e}_z \equiv  \Omega_0 \boldsymbol{n}_{cl} + \Delta \boldsymbol{e}_z,\label{dr20} \\
\Gamma &= \frac{\pi}{2}  J f \sin \phi .
\label{dr23}
\end{align}
The term $\Omega_0 \boldsymbol{n}_{cl}$ in the former equation represents the Lamb shift of the Zeeman field $\Delta \boldsymbol{e}_z$.

The solution of Eq. \eqref{dr19} reads
\begin{align}
\braket{\boldsymbol{S}(t)} &= \braket{\boldsymbol{S}}_{st} +  P_{\boldsymbol{n}_h} [\braket{\boldsymbol{S}}_0 - \braket{\boldsymbol{S}}_{st}] e^{-\Gamma t}
\notag \\
&+  (1 - P_{\boldsymbol{n}_h}) [\braket{\boldsymbol{S}}_0- \braket{\boldsymbol{S}}_{st}] e^{-\Gamma t} \cos \Omega t 
\notag \\
&+   \boldsymbol{n}_{h} \times [\braket{\boldsymbol{S}}_0 - \braket{\boldsymbol{S}}_{st}] e^{-\Gamma t} \sin \Omega t,
\label{dr24} 
\end{align}
where we have introduced 
\beq
\Omega = |\boldsymbol{h}_{eff}|, \qquad \boldsymbol{n}_{h} =\frac{\boldsymbol{h}_{eff}}{\Omega} , 
\eeq
and $P_{\boldsymbol{n}_h}$ is a projector onto the unit vector $\boldsymbol{n}_{h}$.

As one can see from \eqref{dr19}, the Bloch vector of the local spin precesses with the angular frequency $\Omega$ about the axis along the unit vector $ \boldsymbol{n}_h$. All  components of $\boldsymbol{S}(t)$, both longitudinal and transverse, decay with the same decay rate $\Gamma$. Both parameters $\Omega$ and $\Gamma$ essentially depend on the photon density $f$ and hence on the incoming power of the pulse. The precession axis $\boldsymbol{n}_h$ has nontrivial dependence on various parameters only when $\boldsymbol{n}_h$ is noncollinear with $\boldsymbol{e}_z$ (i.e. for any polarization other than purely right or purely left).

The stationary value of the Bloch vector is given by 
\begin{align}
\braket{\boldsymbol{S}}_{st} =  \frac{ \boldsymbol{n}_{cl} + \lambda \boldsymbol{n}_{h} \times \boldsymbol{n}_{cl} + \lambda^2 \cos \psi \, \boldsymbol{n}_{h} }{2 (1 + \lambda^2)},
\label{Sst}
\end{align}
where $\lambda = \Omega/\Gamma$ and $\cos \psi = \boldsymbol{n}_h \cdot \boldsymbol{n}_{cl}$, 
while the initial expectation value $\braket{\boldsymbol{S}}_0 = \braket{\boldsymbol{S}(t = 0)}$ is expressed via initial probability amplitudes of the states $|+ \rangle$ and $| - \rangle$. We assume the absence of initial correlations between photons and the spin, which in particular means that the spin is initially in a pure state. This implies the normalization  $|\braket{\boldsymbol{S}}_0| =\frac12$.

During the time evolution the norm of the Bloch vector changes becoming smaller than $\frac12$, which means that the spin reduced density matrix $\rho_s (t)$ corresponds to a mixed state. It is convenient to characterize the latter by the purity
\begin{align}
\gamma (t) = \mbox{tr} \rho_s^2 (t) = \frac{1 + 4 \braket{{\bf S}(t)}^2}{2},
\label{dr30a}
\end{align}
which in general takes values in the range between $\frac12$ (maximally mixed state with the maximal entropy $\ln 2$) and $1$ (pure state, zero entropy).

\begin{figure}[ht]
 \includegraphics[width=90mm]{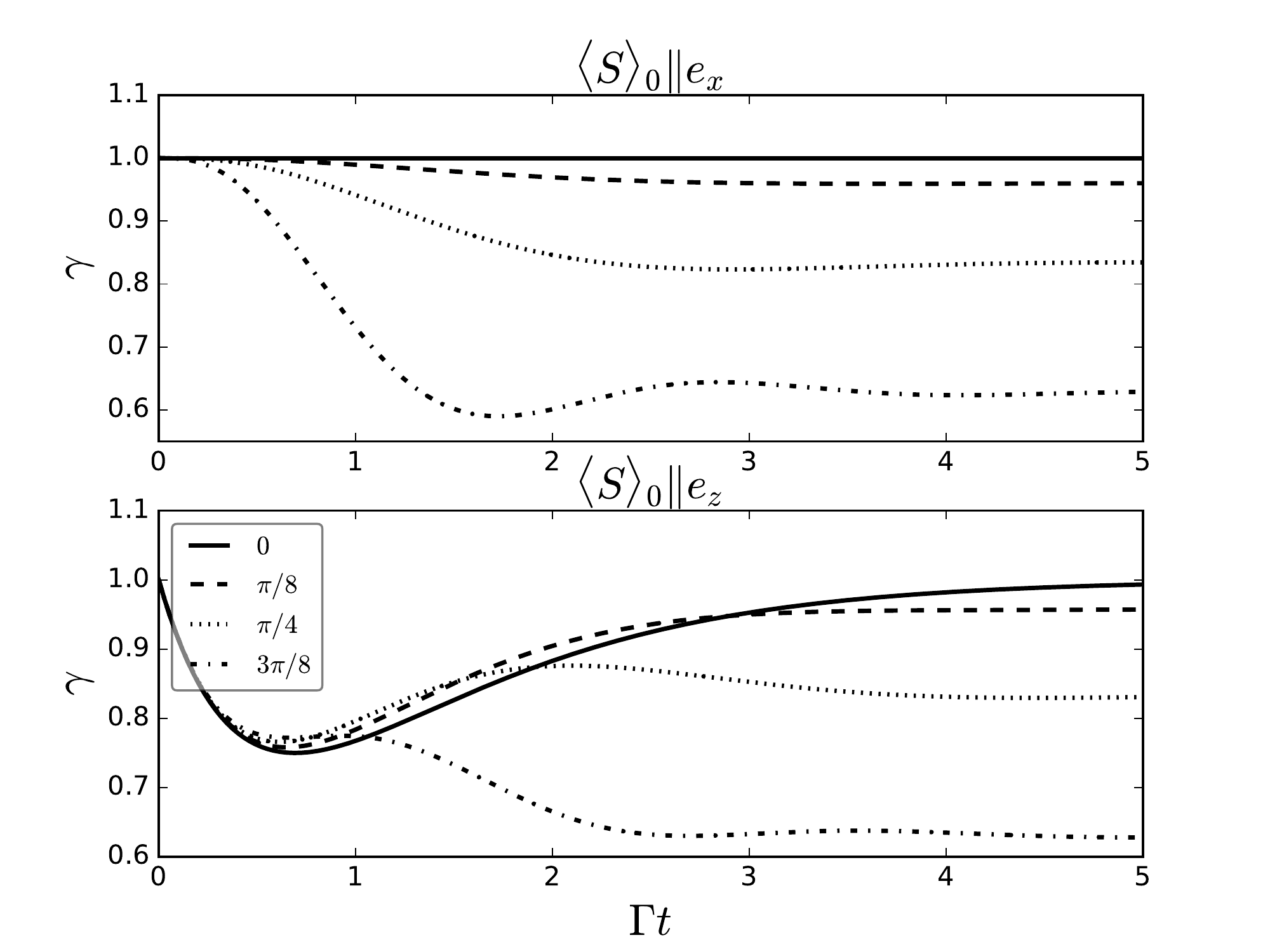}\\
    \caption{ Purity $\gamma (t)$ of the local two-level system for different values of $\psi = \arccos (\boldsymbol{n}_{cl} \cdot\boldsymbol{n}_{h})$. For $\Delta \neq 0$ the stationary value of the purity is less than unity, while for $\Delta = 0$ the purity approaches unity (pure state) in the long time limit. Initial conditions are determined  by the two different alignments, $\braket{\boldsymbol{S}}_0   \parallel \boldsymbol{e}_{x} = \boldsymbol{n}_{cl}$ and $\braket{\boldsymbol{S}}_0 \parallel \boldsymbol{e}_{z} \perp \boldsymbol{n}_{cl}$. Stationary values of the purity do not depend on an initial state.} \label{purity}
\end{figure}

In the stationary limit $t \rightarrow \infty$ we find
\begin{align}
\gamma_{st} =  \frac{1 + \frac14 \lambda^2 (3 + \cos 2 \psi)}{1 + \lambda^2}.
\label{dr30b}
\end{align}
The pure state occurs for $\psi =0$ and $\psi =\pi$. There are two cases when this state can be reached.
\begin{enumerate}
\item If $\Delta = 0$, we find $\boldsymbol{n}_{h} = \boldsymbol{n}_{cl}$, and therefore $\psi = 0$. This case corresponds to a degenerate two-level system which obviously cannot change energy of incoming photons (in the long time limit) and hence it cannot mediate inelastic processes which might lead to an entanglement between the field and the spin. Thus  the state of these two subsystems in the far future after the interaction will be again factorized, and the purity of the spin state reaches unity. This observation confirms our argument stated in Section \ref{Consider}.
\item If either $\alpha_+$ or $\alpha_-$ vanishes,  we also find $\boldsymbol{n}_{cl} = \pm \boldsymbol{e}_z$ and $\boldsymbol{n}_{h} = \pm \boldsymbol{e}_z$, and thus $\psi = 0$ or $\psi =\pi$. This case corresponds to a (right or left) circularly polarized  incident beam of photons. Suppose that we have the right polarized beam (the Jones vector $\boldsymbol{s}_{cl} = \frac{f}{2} \boldsymbol{e}_z$). Since the spontaneous emission is proportional to the mean number of photons driving the corresponding transition $|+ \rangle \leftrightarrow |\tilde{+}\rangle$, then the local system will end up in the state $|+\rangle$, corresponding to the Bloch vector $\frac{1}{2} \boldsymbol{e}_z$. Thus there are no inelastic processes and, hence, no entanglement in the stationary limit.
\end{enumerate}
If neither of these conditions is met, that is if the field is elliptically polarized and the two-level splitting $\Delta$ (= the  two-photon detuning of the original model) is finite, the field and the local system will finally end up in an entangled state. The resulting field is not coherent anymore, and therefore the field-field correlation functions do not factorize. This is accompanied by inelastic scattering of photons into modes different from the input one, $\omega_0$.

In Fig. \ref{purity} we show time dependence of the purity for different alignments (parallel and perpendicular) of the initial Jones vector and the initial Bloch vector at various values of the two-photon detuning $\Delta$.

Having understood time dynamics of the local spin and its stationary values, we go over to a description of the field properties in the stationary limit.

\section{First order correlation functions and power spectrum}

Power  spectra  of  three-level  atoms  were  discussed  in several papers over the years, see \cite{CT-R}, \cite{WS}, \cite{Sobol}, \cite{Fu},  \cite{Mavroyannis1}, \cite{Mavroyannis2}, \cite{AB}, \cite{BPK}, \cite{exp-qd}.  However none of these studies has dealt with polarization effects in the excitation spectrum. Here we fill up this gap  by making exact calculations of the polarization-dependent power spectrum.

\subsection{Polarization unresolved power spectrum}
\label{sub3A}

Using Eqs. \eqref{dr14} and \eqref{Sst} we can express  the stationary value of the polarization unresolved field-field correlation function by
\begin{align}
C^0 (\tau) &= \sum_{\sigma} \braket{a^{\dagger}_{\sigma}(x,t + \tau) a_{\sigma}(x, t)}  \label{dr32}  \\
&=  f e^{i \omega_0 \tau}  [ 1 -  \left( \frac{5}{4} - \gamma_{st} \right) \sin^2 \frac{\phi}{2}   \notag\\
&   +    \braket{\boldsymbol{S}(t - x+ \tau) \cdot \boldsymbol{S}(t - x)} \sin^2 \frac{\phi}{2}  \notag  \\
&  + i  {\boldsymbol{n}}_{cl} \cdot \braket{\boldsymbol{S}(t - x + \tau) \times \boldsymbol{S}(t - x)} \sin^2 \frac{\phi}{2} ].
 \notag
\end{align}
In this derivation we also used the identity
\begin{align}
\lbrack a_{\sigma}(-t - \tau,0) , S^i (t) \rbrack &= 0 \hspace{0.5cm} \mathrm{for} \quad \tau > 0,
\label{dr34}
\end{align} 
which follows from the commutation of the Heisenberg operators $\lbrack a_{\omega \sigma}(t) , S^i (t) \rbrack = 0$ (see Appendix \ref{App:AppendixC} for the proof). The relation \eqref{dr34} is a formal expression of the causality principle, stating that the free field at the position $x= - \tau<0$ and time $t$ (that is, to the left from the scatterer, and hence before interacting with it), which is described by the Schr\"odinger operator $a_{\sigma}(-t - \tau,0)$, remains independent of the scatterer's state  $S^i (t)$.

We can decompose \eqref{dr32} into elastic (or "coherent") and inelastic (or "incoherent") contributions
\begin{align}
C^0(\tau) &= C_{el}^0(\tau)+ C_{inel}^0(\tau), \\
C_{el}^0(\tau)
&=  f e^{i \omega_0 \tau}  \left[ 1 -  \frac32\left(1 - \gamma_{st} \right) \sin^2 \frac{\phi}{2}   \right] , \\
C_{inel}^0(\tau)
&=  f e^{i \omega_0 \tau} \sin^2  \frac{\phi}{2}  [   \braket{\boldsymbol{S}(t+ \tau) \cdot \boldsymbol{S}(t)}  - \braket{\boldsymbol{S}}_{st}^2  \notag  \\
&  \qquad \qquad + i  {\boldsymbol{n}}_{cl} \cdot \braket{\boldsymbol{S}(t + \tau) \times \boldsymbol{S}(t)}  ].
\end{align}

\begin{figure}[t]
\centering
\includegraphics[width=90mm]{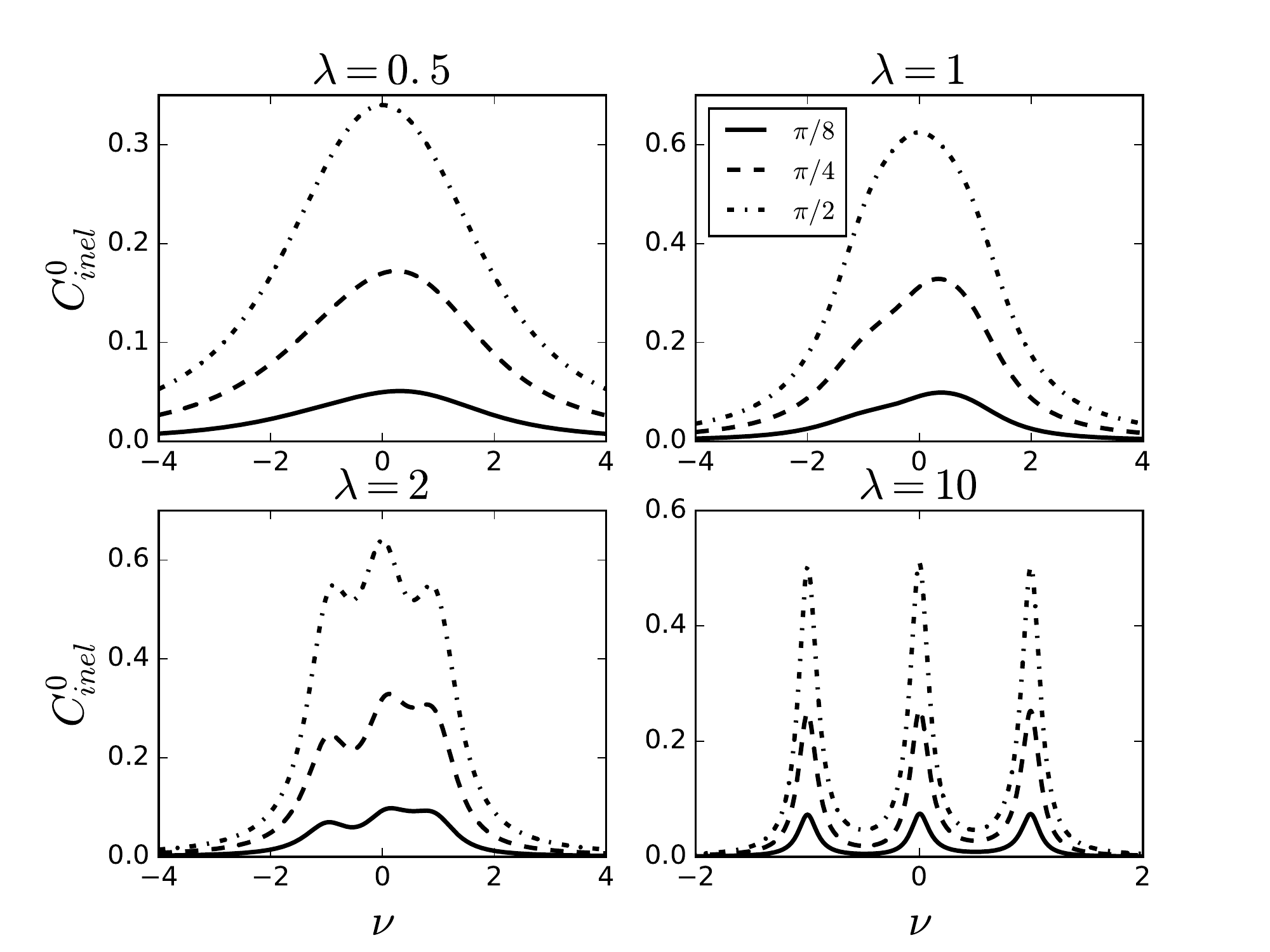}
\caption{Inelastic peaks in the polarization unresolved power spectra, Eq. \eqref{PS}. The legend in the upper right panel indicates the values of $\psi$, which  apply to all panels. For large  values of $\Delta \gg J f$ (which implies $\lambda \gg 1$) the peaks are well resolved, and one can extract the values of $\Omega$ and $\Gamma$.}
\label{unres_spec}
\end{figure}

For $\tau > 0$ the equation of motion for $C^{ij}(\tau)$ in the long time limit reads
\begin{align}
\frac{d}{d\tau} C^{ij}(\tau) = \Omega \epsilon_{ilk} n^l_{h} C^{kj}(\tau) - \Gamma \left[ C^{ij}(\tau) - \frac{n^i_{cl} \braket{S^j}_{st}}{2} \right].
\label{dr34a}
\end{align}
Solving these equations we can compute \eqref{dr32} for $\tau \geq 0$. The Fourier transform of \eqref{dr32} with respect to $\tau$ yields the sought power spectrum. Providing the details of this calculation in Appendix \ref{App:AppendixC}, we present
the polarization unresolved power spectrum
\begin{align}
& \tilde{C}^0 (\nu) = \frac{2 \pi f}{\Omega} \delta(\nu) \left[1 - \frac{3}{2} (1 - \gamma_{st}) \sin^2 \frac{\phi}{2} \right]
\label{ps79} \\
\begin{split}
&+ \frac{f (1 - \gamma_{st})}{\Gamma}  \sin^2 \frac{\phi}{2}  \left[  \frac{1 + \nu \cos \psi }{1 + \lambda^2 \nu^2}  \right.  \\
&+\left. \frac{1 - (\nu-1)  \cos^2 \frac{\psi}{2}}{1+ \lambda^2 (\nu-1)^2} + \frac{1 + (\nu + 1) \sin^2 \frac{\psi}{2} }{1+ \lambda^2 (\nu +1)^2}\right] ,
\end{split}
\label{PS}
\end{align}
where $\nu =(\omega - \omega_0)/\Omega$.
Note that the inelastic power spectrum is proportional to  $(1 - \gamma_{st})$, and it vanishes for the pure stationary state of the local system.

Fig. \ref{unres_spec} shows $\tilde{C}^0_{inel}(\nu)$ for various values of $\lambda$ and $\psi$.
It exhibits the three-peak structure, as suggested by the heuristic arguments made in Section \ref{Consider}. The resonances become sharper for increasing $\lambda$, which is achieved at a weak coupling $J$ and a weak field $f$ compared against $\Delta$. The Stokes shift $\Omega = |\boldsymbol{h}_{eff}|$, given by the magnitude of an effective "magnetic field" experienced by the local system, can be extracted from this plot along with the peaks' broadenings $\Gamma$.  

The shape of $\tilde{C}^0_{inel}(\nu)$ is invariant under simultaneous transformations $\nu \to - \nu$ and $\psi \rightarrow \psi + \frac{\pi}{2}$, which correspond to a sign change of $\Delta$ and an exchange of roles between the left and the right polarizations. Under the  transformation $\nu \to - \nu$ alone, the power spectrum $\tilde{C}^0_{inel}(\nu)$ is not invariant, that is asymmetric, which means that the rates of the processes (i) and (ii) defined in Section \ref{Consider} are in general different and dependent on the initial polarization.

The integrated output power amounts to 
\begin{align}
P_{tot} = \frac{\Omega}{2 \pi} \int d \nu (\nu \Omega + \omega_0) \tilde{C}^0 (\nu) = \omega_0 f,
\label{dr36}
\end{align}
and thereby it is conserved. The inelastically scattered contribution to it is given by 
\begin{align}
P_{inel} = \frac{3 \omega_0  f }{2} (1- \gamma_{st }) \sin^2 \frac{\phi}{2} .
\end{align}

\subsection{Polarization resolved spectral function}
To probe cross-correlations between different polarizations it is useful to define the polarization resolved first order correlation function 
\begin{align}
C^d (\tau) & = \sum_{\sigma, \sigma'} \braket{a^{\dagger}_{\sigma}(x,t + \tau) \frac{\delta_{\sigma \sigma^{\prime}} + n^i_d \sigma^i_{\sigma \sigma^{\prime}}}{2} a_{\sigma^{\prime}}(x,t)} \nonumber\\
&= C^0(\tau) + \boldsymbol{n}_d \cdot \boldsymbol{C}_s(\tau),
\label{dr37}
\end{align}
which resolves photonic polarizations parallel to  $\pm \boldsymbol{n}_d$. The unit vector $\boldsymbol{n}_d$ thus defines a detector's polarization. 

Using Eq. \eqref{dr14} we express $\boldsymbol{C}_s(\tau)$ in the long time limit  as a sum of elastic and inelastic contributions
\begin{align}
& \boldsymbol{C}_s(\tau)  =  \boldsymbol{C}_{s,el} (\tau)  + \boldsymbol{C}_{s,inel} (\tau) , \label{dr38} \\
& \boldsymbol{C}_{s,el} (\tau) =  \frac{f}{2} e^{i \omega_0 \tau}[  \cos^2 \frac{\phi}{2} \boldsymbol{n}_{cl} +2 \sin^2 \frac{\phi}{2}  \braket{\boldsymbol{S}}_{st} \notag \\
&  \qquad \qquad -  \sin \phi \, \boldsymbol{n}_{cl} \times \braket{\boldsymbol{S}}_{st} ]  \notag \\
& \qquad + \frac{f}{4} e^{i \omega_0 \tau}  \sin^2 \frac{\phi}{2} (1- \gamma_{st}) (  \boldsymbol{n}_{cl} -  4 \braket{\boldsymbol{S}}_{st}), \\
& \boldsymbol{C}_{s,inel} (\tau) = \frac{f}{2} e^{i \omega_0 \tau}  \sin^2 \frac{\phi}{2} \notag \\
 & \times [ \langle \boldsymbol{S}(t + \tau )  (\boldsymbol{n}_{cl} \cdot \boldsymbol{S}(t)) \rangle + \braket{(\boldsymbol{n}_{cl} \cdot \boldsymbol{S}(t + \tau )) \boldsymbol{S}(t)}  \notag \\
&  \qquad \qquad -2 ( \boldsymbol{n}_{cl} \cdot \braket{\boldsymbol{S}}_{st} ) \braket{\boldsymbol{S}}_{st}  \notag \\
& \qquad  -  \boldsymbol{n}_{cl}( \braket{\boldsymbol{S}(t + \tau ) \cdot \boldsymbol{S}(t )} - \braket{\boldsymbol{S}}_{st}^2 ) \nonumber\\
&  \qquad - i \braket{\boldsymbol{S}(t + \tau ) \times \boldsymbol{S}(t )}],
\label{dr40}
\end{align}
where the inelastic contribution is expressed via the spin-spin correlation functions.

From \eqref{ps8} we get at $t,x \to \infty$ the stationary value of the average field
\begin{align}
\langle a_{\sigma}(x,t) \rangle_{st} &= \frac{e^{i \omega_0 (x-t)}}{\sqrt{L}}  \sum_{\sigma'} U_{\sigma \sigma'}  \alpha_{\sigma'} ,
 \notag \\
U_{\sigma \sigma'} & =\frac{1 - e^{i \phi}}{2} \sigma_{\sigma \sigma'}^i S_{st}^i + \frac{3 + e^{i \phi}}{4} \delta_{\sigma \sigma'}.
\end{align}
Its polarization is parameterized by the new Jones vector 
\begin{align}
\boldsymbol{s}_q & \equiv \sum_{\sigma , \sigma'} \langle a_{\sigma}^{\dagger} (x,t) \rangle_{st} \frac{\boldsymbol{\sigma}_{\sigma \sigma'}}{2}  \langle a_{\sigma'}(x,t) \rangle_{st}  = \boldsymbol{C}_{s,el} (0) .
\end{align}
We label it by the subscript $q$ to indicate that the initially classical state is affected by interaction with the quantum system. In the absence of interaction ($\phi=0$) we recover $\boldsymbol{s}_q = \boldsymbol{s}_{cl}$.

\begin{figure}[b]
\centering
\includegraphics[width=90mm]{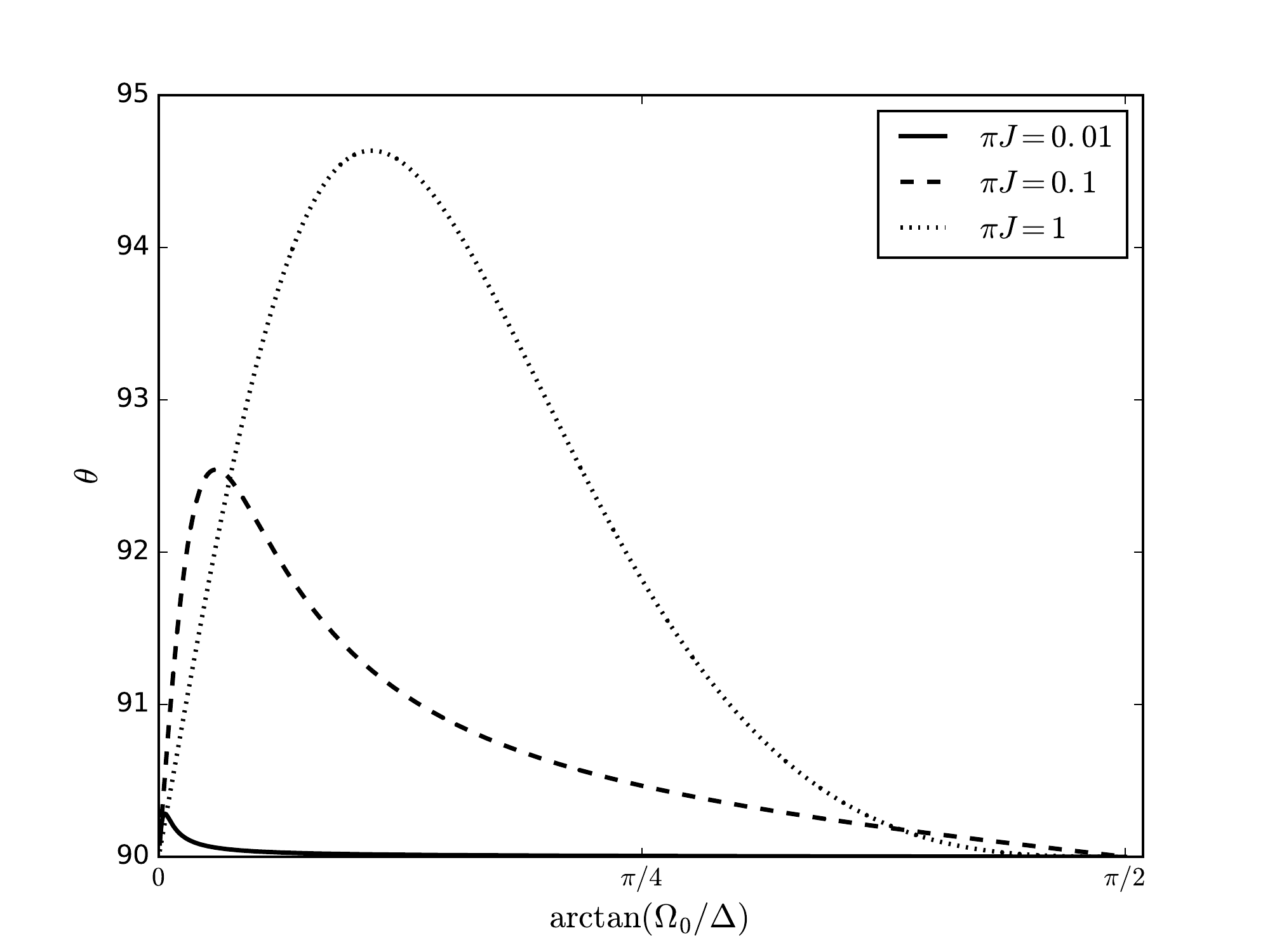}
\caption{Ellipticity $\theta = \arccos  \frac{\boldsymbol{e}_z \cdot \boldsymbol{s}_q}{|\boldsymbol{s}_q|}$ (in degrees) as a function of the ratio between the two-photon detuning and its Lamb shift at various exchange couplings $J$. An initially linearly polarized light with $\boldsymbol{n}_{cl} = \boldsymbol{e}_x$ does not considerably change its type of polarization: deviations from the equator on the Jones sphere do not exceed five degrees.}
\label{ellip}
\end{figure}

In Fig. \ref{ellip} we plot the angle $\theta$ parameterizing a degree of ellipticity in a polarization of the emitted radiation, provided that initially it was linear, $\boldsymbol{n}_{cl} = \boldsymbol{e}_x$. Although a type of polarization changes very slightly from the incoming to the outgoing pulse, one can experimentally extract the quantity $\frac{\Delta}{\pi J \cos^2 \frac{\phi}{2}}$ by measuring a polarization of the average emitted field at various powers of the incoming signal $\propto f$ and fitting the result to the curves in Fig. \ref{ellip}. In other words, by polarization measurements one can find the ratio $\Omega_0 / \Delta$ for given $f$. Combining this result with the ratio $\Omega / \Gamma$ extracted from the power spectra shown in Fig. \ref{unres_spec}, it becomes possible to find the values of $\Delta / \Gamma$ and $\Omega_0 / \Gamma$, characterizing the local system.

Analogously to the unresolved case we compute the spin-spin correlators in Eq. \eqref{dr40}  (see Appendix \ref{App:corr-res} for details). Performing the Fourier transform of  \eqref{pr40} we find the direction dependent part of the polarization resolved inelastic power spectrum
\begin{widetext}
\begin{align}
\tilde{\boldsymbol{C}}_{s,inel} (\nu) & = \frac{f}{2 \Omega} \sin^2 \frac{\phi}{2} (1- \gamma_{st}) \notag \\
\times & \left\{ -  \frac{\lambda  \boldsymbol{n}_{cl} - 2 \lambda  \cos \psi  \boldsymbol{n}_{h} + \boldsymbol{n}_{cl} \times \boldsymbol{n}_h}{\lambda^2 \nu^2 +1}  + \frac{\lambda \nu \boldsymbol{n}_{h} }{\lambda^2 \nu^2 +1} \right.  \notag  \\
&+  \frac{ 2 \lambda  \boldsymbol{n}_{cl}+   2  \lambda  (1+ \lambda^2 - \cos \psi )  \boldsymbol{n}_{h} + (1 - \lambda^2) \boldsymbol{n}_{cl} \times \boldsymbol{n}_h}{2 (1+ \lambda^2) [\lambda^2 (\nu - 1)^2 +1]} \notag \\
& - \lambda (\nu -1)  \frac{  (1 - \lambda^2) \boldsymbol{n}_{cl}+ (1+ \lambda^2 + 2 \lambda^2\cos \psi ) \boldsymbol{n}_{h} -  2  \lambda\boldsymbol{n}_{cl} \times \boldsymbol{n}_h}{2 (1+ \lambda^2) [\lambda^2 (\nu - 1)^2 +1]} \notag \\
&+   \frac{ 2 \lambda  \boldsymbol{n}_{cl} -  2  \lambda  (1+ \lambda^2 + \cos \psi )  \boldsymbol{n}_{h} + ( 1 - \lambda^2) \boldsymbol{n}_{cl} \times \boldsymbol{n}_h}{2 (1+ \lambda^2) [\lambda^2 (\nu + 1)^2 +1] }   \notag \\
& \left. + \lambda (\nu + 1)  \frac{(1 - \lambda^2)  \boldsymbol{n}_{cl} -   ( 1+ \lambda^2 - 2 \lambda^2 \cos \psi ) \boldsymbol{n}_{h}   - 2  \lambda \boldsymbol{n}_{cl} \times \boldsymbol{n}_h}{2 (1+ \lambda^2) [\lambda^2 (\nu + 1)^2 +1] } \right\} . 
\label{dr41}
\end{align}
\end{widetext}

\begin{figure}
\centering
\includegraphics[width=90mm]{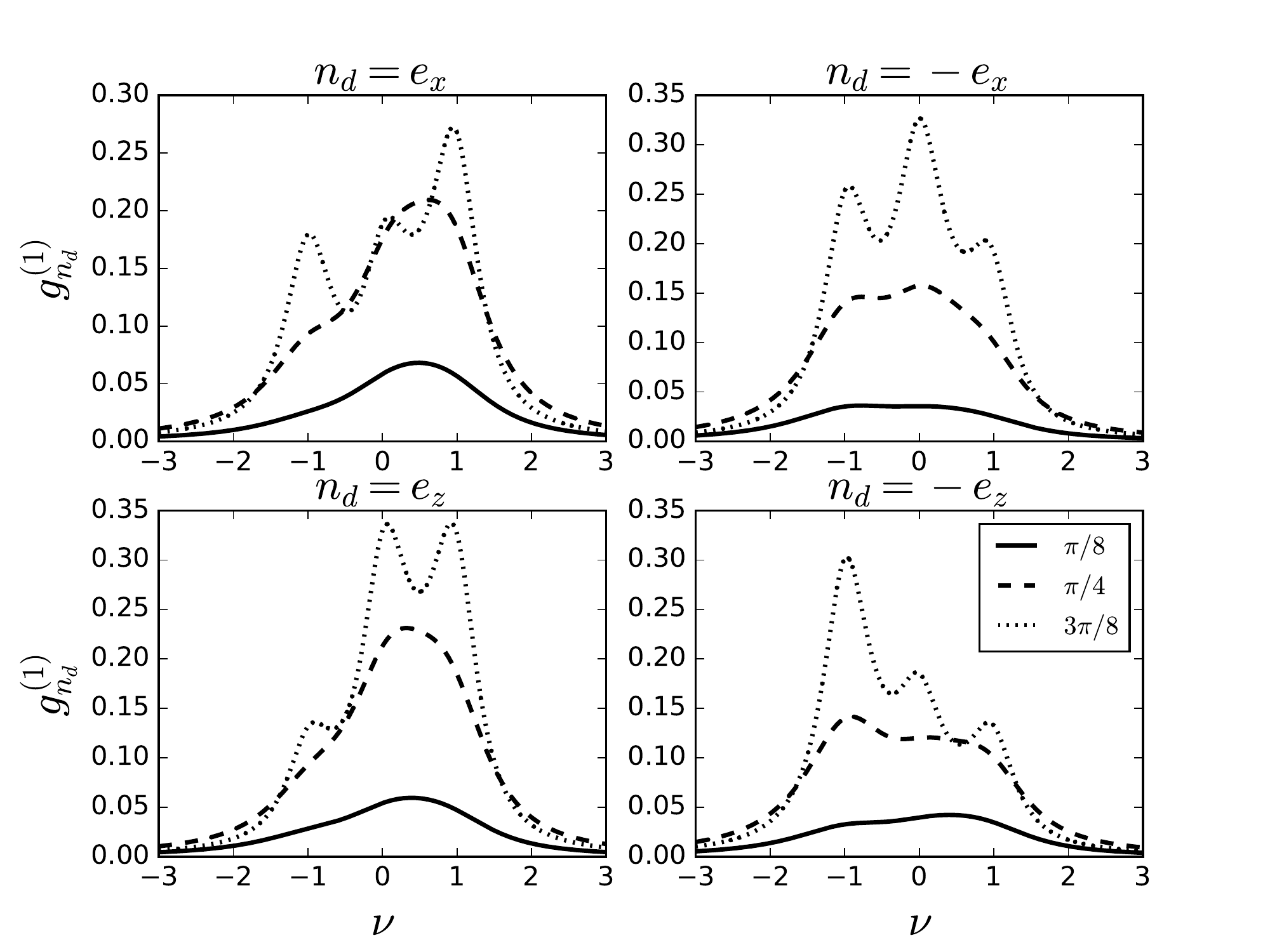}
\caption{Polarization resolved
inelastic power spectrum for various values of $\psi$ (shown in the legend) and detector polarizations $\boldsymbol{n}_d$. In particular, the two lower panels show "how many" photons with right and left circular polarizations can be produced from the initial pulse linearly polarized along  $\boldsymbol{e}_x$.}
\label{res_spec}
\end{figure}

In Fig. \ref{res_spec} we plot  $g_{\boldsymbol{n}_d}^{(1)} (\nu) = \tilde{C}^0 (\nu) + \boldsymbol{n}_d  \cdot \tilde{\boldsymbol{C}}_{s,inel} (\nu)$. A detector selects photons with the polarization parameterized by the unit vector $\boldsymbol{n}_d$, and the polarization resolved inelastic power spectrum $g_{\boldsymbol{n}_d}^{(1)} (\nu)$ contains information about the weights of frequency modes to which these photons scatter. For the initial linear polarization along $\boldsymbol{e}_x$ we choose $\boldsymbol{n}_d = \pm \boldsymbol{e}_x$ (upper panel) and $\boldsymbol{n}_d = \pm \boldsymbol{e}_z$ (lower panels). The last two cases show the mean number of photons with right and left circular polarizations in different frequency modes which are produced  by the local system from the linearly polarized input pulse.

We note that the effect of the polarization change is much more pronounced and detailed in the function $g_{\boldsymbol{n}_d}^{(1)} (\nu)$  than in the polarization of the average field depicted in Fig. \ref{ellip}. In addition, we emphasize that  the effective "magnetic field" $\boldsymbol{h}_{eff}$ produced by the local system breaks the symmetry between the right and left circularly polarized photons, since the vector $\boldsymbol{n}_h$ necessarily lies in the $x - z$ plane.

\section{Second order correlation function}
\label{g2sect}

Statistical properties of the emitted light are characterized by the second order correlation functions 
\begin{align}
& G^{(2)}_{{\sigma}'_1  {\sigma}'_2 , \sigma_1 \sigma_2 } \notag \\
&= \braket{a^{\dagger}_{{\sigma}^{\prime}_1}(x,t) a^{\dagger}_{\sigma^{\prime}_2}(x,t + \tau) a_{\sigma_2} (x,t + \tau) a_{\sigma_1} (x,t) }.
\label{dr46}
\end{align}
It is more convenient to introduce the following combinations
\begin{align}
G_{\boldsymbol{n},\boldsymbol{m}}(\tau)&= \frac14 \sum_{\sigma'_2 ,\sigma_2} \sum_{\sigma'_1 ,\sigma_1} (\boldsymbol{n} \cdot\boldsymbol{\sigma}_{\sigma'_2 \sigma_2})(\boldsymbol{m} \cdot \boldsymbol{\sigma}_{\sigma'_1 \sigma_1} )G^{(2)}_{\sigma_{1}'\sigma_{2}' , \sigma_{1}\sigma_{2}}, \label{Gnm}
\end{align}
\begin{align}
G_{\boldsymbol{n},0}(\tau) &= \frac12 \sum_{\sigma'_2 ,\sigma_2} \sum_{\sigma_1} (\boldsymbol{n} \cdot \boldsymbol{\sigma}_{\sigma^{\prime}_2 \sigma_2})G^{(2)}_{\sigma_{1}\sigma_{2}' , \sigma_{1}\sigma_{2}} , \label{Gn0}
\end{align}
\begin{align}
G_{0,\boldsymbol{m}}(\tau) &= \frac12 \sum_{\sigma_2}   \sum_{\sigma'_1, \sigma_1}  (\boldsymbol{m} \cdot \boldsymbol{\sigma}_{\sigma^{\prime}_1 \sigma_1})G^{(2)}_{\sigma_{1}' \sigma_{2}, \sigma_{1}\sigma_{2}} , \label{G0n} \\
G_{0,0}(\tau)&=   \sum_{\sigma_2}   \sum_{\sigma_1}  G^{(2)}_{\sigma_{1}\sigma_{2}, \sigma_{1}\sigma_{2}} , \label{G00}
\end{align}
and consequently 
\begin{align}
& g^{(2)}_{\boldsymbol{n},\boldsymbol{m}}  (\tau) = \left[ G_{\boldsymbol{n},\boldsymbol{m}} (\tau)+ \frac{G_{0,\boldsymbol{m}} (\tau)+ G_{\boldsymbol{n},0} (\tau)}{2}+ \frac{G_{0,0} (\tau)}{4}\right]     \notag \\
& \times \left[ G_{\boldsymbol{n},\boldsymbol{m}} (\infty)+ \frac{G_{0,\boldsymbol{m}} (\infty)+ G_{\boldsymbol{n},0} (\infty)}{2}+ \frac{G_{0,0} (\infty)}{4}\right]^{-1}  .
\end{align}
Physical meaning of the latter function is transparent: the second (first) index indicates the polarization vector of the first (second) measured photon in coincidence measurements with delay time $\tau$.

To find the quantities \eqref{Gnm}-\eqref{G00} we use the relations directly following from \eqref{dr14}
\begin{align}
&\sum_{\sigma', \sigma} a^{\dagger}_{\sigma'}(x,t) \frac{\boldsymbol{\sigma}_{\sigma' \sigma}}{2} a_{\sigma}(x,t)
\notag \\
& =  \sum_{\sigma', \sigma}  a^{\dagger}_{\sigma'}(x-t,0)  \frac12  [  \cos^2 \frac{\phi}{2} \boldsymbol{\sigma}_{\sigma' \sigma} + \sin^2 \frac{\phi}{2} \delta_{\sigma' \sigma} \boldsymbol{S}(t-x) \nonumber\\
&  - \sin \phi \,\, \boldsymbol{\sigma}_{\sigma' \sigma} \times \boldsymbol{S}(t-x) ] a_{\sigma}(x-t,0),
\label{gt5} \\
& \sum_{\sigma} a^{\dagger}_{\sigma}(x,t) a_{\sigma}(x,t) = \sum_{\sigma}  a^{\dagger}_{\sigma}(x-t,0) a_{\sigma}(x-t,0).
\end{align}
They allow us to find (see Appendix \ref{App:g2} for details) that in the stationary limit
\begin{align}
G_{\boldsymbol{n},\boldsymbol{m}}(\tau)&= \frac{f}{2} (\boldsymbol{m} \cdot \boldsymbol{C}_{s} (0)) \, \boldsymbol{n} \cdot \!  [ \cos^2 \frac{\phi}{2}  \boldsymbol{n}_{cl} + 2 \sin^2 \frac{\phi}{2} \boldsymbol{k}_{\boldsymbol{m}}(\tau)  \notag \\
&     - \sin \phi \,  \boldsymbol{n}_{cl} \times \boldsymbol{k}_{\boldsymbol{m}}(\tau) ] ,  \\
G_{\boldsymbol{n},0}(\tau)&=  \frac{f^2}{2}  \boldsymbol{n} \cdot  [ \cos^2 \frac{\phi}{2} \boldsymbol{n}_{cl} +2 \sin^2 \frac{\phi}{2} \boldsymbol{k}_{0}(\tau)  \notag \\
&   - \sin \phi \,  \boldsymbol{n}_{cl} \times \boldsymbol{k}_{0}(\tau) ], \\
G_{0,\boldsymbol{m}}(\tau)&=  f  (\boldsymbol{m} \cdot \boldsymbol{C}_{s} (0)) , \\
G_{0,0}(\tau) &= f^2 ,
\end{align}
where 
\beq
\boldsymbol{C}_s (0) &= & \frac{f}{2}  [ \cos^2 \frac{\phi}{2} \boldsymbol{n}_{cl} + 2 \sin^2 \frac{\phi}{2} \braket{\boldsymbol{S}}_{st}  \nonumber\\
& &  -  \sin \phi \, \boldsymbol{n}_{cl} \times \braket{\boldsymbol{S}}_{st} ] ,
\label{dr40a}
\eeq
and the vectors $\boldsymbol{k}_{0}(\tau)$ and $\boldsymbol{k}_{\boldsymbol{m}}(\tau)$
\begin{align}
\frac{d}{d\tau} \boldsymbol{k}_{0}(\tau) &= \Omega \boldsymbol{n}_h \times \boldsymbol{k}_{0}(\tau) - \Gamma [ \boldsymbol{k}_{0}(\tau) - \frac12 \boldsymbol{n}_{cl}  ],
\label{dr48} \\
\frac{d}{d\tau} \boldsymbol{k}_{\boldsymbol{m}}(\tau) &= \Omega \boldsymbol{n}_h \times \boldsymbol{k}_{\boldsymbol{m}}(\tau) - \Gamma [ \boldsymbol{k}_{\boldsymbol{m}}(\tau) - \frac{1}{2} \boldsymbol{n}_{cl} ],
\label{gt16a} 
\end{align}
with initial condition given by 
\begin{align}
\boldsymbol{k}_0 (0) &=  \braket{\boldsymbol{S}}_{st} \cos^2 \frac{\phi}{2} + \frac{1}{2} \boldsymbol{n}_{cl} \sin^2 \frac{\phi}{2} \notag \\
&+ \frac{1}{2} \boldsymbol{n}_{cl} \times \braket{\boldsymbol{S}}_{st} \sin \phi, 
\label{dr49} \\
\boldsymbol{k}_{\boldsymbol{m}}(0) &= \frac{f}{2 (\boldsymbol{C}_{s} (0) \cdot\boldsymbol{m})} \left[ (\boldsymbol{n}_{cl}\cdot\boldsymbol{m}) \braket{\boldsymbol{S}}_{st} \cos^2 \frac{\phi}{2}  \right. \notag \\
& +  {\boldsymbol{n}}_{cl} (\braket{\boldsymbol{S}}_{st}\cdot \boldsymbol{m}) \sin^2 \frac{\phi}{2} \notag \\
& \left. -  \frac{\sin \phi}{4} \boldsymbol{m} \times (\boldsymbol{n}_{cl}-2 \braket{\boldsymbol{S}}_{st}) \right].
\end{align}

It is remarkable that the major ingredients $\boldsymbol{k}_{0}(\tau)$ and $\boldsymbol{k}_{\boldsymbol{m}}(\tau)$ of the second order correlation function obey the same equations as the local spin (compare Eqs. \eqref{dr48} and \eqref{gt16a} to Eq. \eqref{dr19}). On the other hand, none of the equations for the first order correlation function (either polarization unresolved or resolved) reminds the equation for the local spin. At first glance this might seem to be a violation of the {\it quantum regression theorem} (see, e.g., \cite{GardinerZoller} for its formulation). However, a deeper analysis shows that in the standard formulation of this theorem the linearity of light-matter interaction in the field operators is required. In our case, this interaction is bilinear in fields, and therefore it becomes necessary to promote the quantum regression theorem to the level of second order correlation functions, what we indeed see in the form of Eqs. \eqref{dr48} and \eqref{gt16a}.

It is now straightforward to find solutions of these equations. Analogously to Eq.~\eqref{dr24} they read
\beq
\boldsymbol{k} (\tau) &=& \braket{\boldsymbol{S}}_{st} +  P_{\boldsymbol{n}_h} [ \boldsymbol{k} (0) - \braket{\boldsymbol{S}}_{st}]  e^{-\Gamma \tau}
\notag \\
&+ & (1 - P_{\boldsymbol{n}_h}) [ \boldsymbol{k} (0) - \braket{\boldsymbol{S}}_{st}] e^{-\Gamma \tau} \cos  \Omega \tau  \notag \\
&+&  \boldsymbol{n}_{h} \times [ \boldsymbol{k}(0) - \braket{\boldsymbol{S}}_{st}] e^{-\Gamma \tau} \sin \Omega \tau , 
\eeq
where $\boldsymbol{k} $ is either $\boldsymbol{k} _0$ or $\boldsymbol{k}_{\boldsymbol{m} }$.

Finally, we obtain
\begin{align}
& g^{(2)}_{\boldsymbol{n},\boldsymbol{m}}  (\tau) = 1+ f \sin \frac{\phi}{2} \nonumber \\
& \times \left[   (\boldsymbol{m} \cdot \boldsymbol{C}_{s} (0)) (\boldsymbol{n} \cdot  \boldsymbol{l}_{\boldsymbol{m}} (\tau))+  \frac{f}{2}  (\boldsymbol{n} \cdot  \boldsymbol{l}_0 (\tau))  \right]\notag\\
&\times \left[ \frac{f}{2} +\boldsymbol{m} \cdot \boldsymbol{C}_{s} (0)\right]^{-1} \left[ \frac{f}{2} +  \boldsymbol{n} \cdot  \boldsymbol{C}_{s} (0) \right]^{-1} ,
\end{align}
where
\begin{align}
\boldsymbol{l} (\tau) & = \sin \frac{\phi}{2} [\boldsymbol{k} (\tau) -\braket{\boldsymbol{S}}_{st}] \notag \\
&  - \cos  \frac{\phi}{2} \boldsymbol{n}_{cl}  \times [\boldsymbol{k} (\tau) -\braket{\boldsymbol{S}}_{st}] . 
\end{align}

\begin{figure}[t]
\centering
\includegraphics[width=90mm]{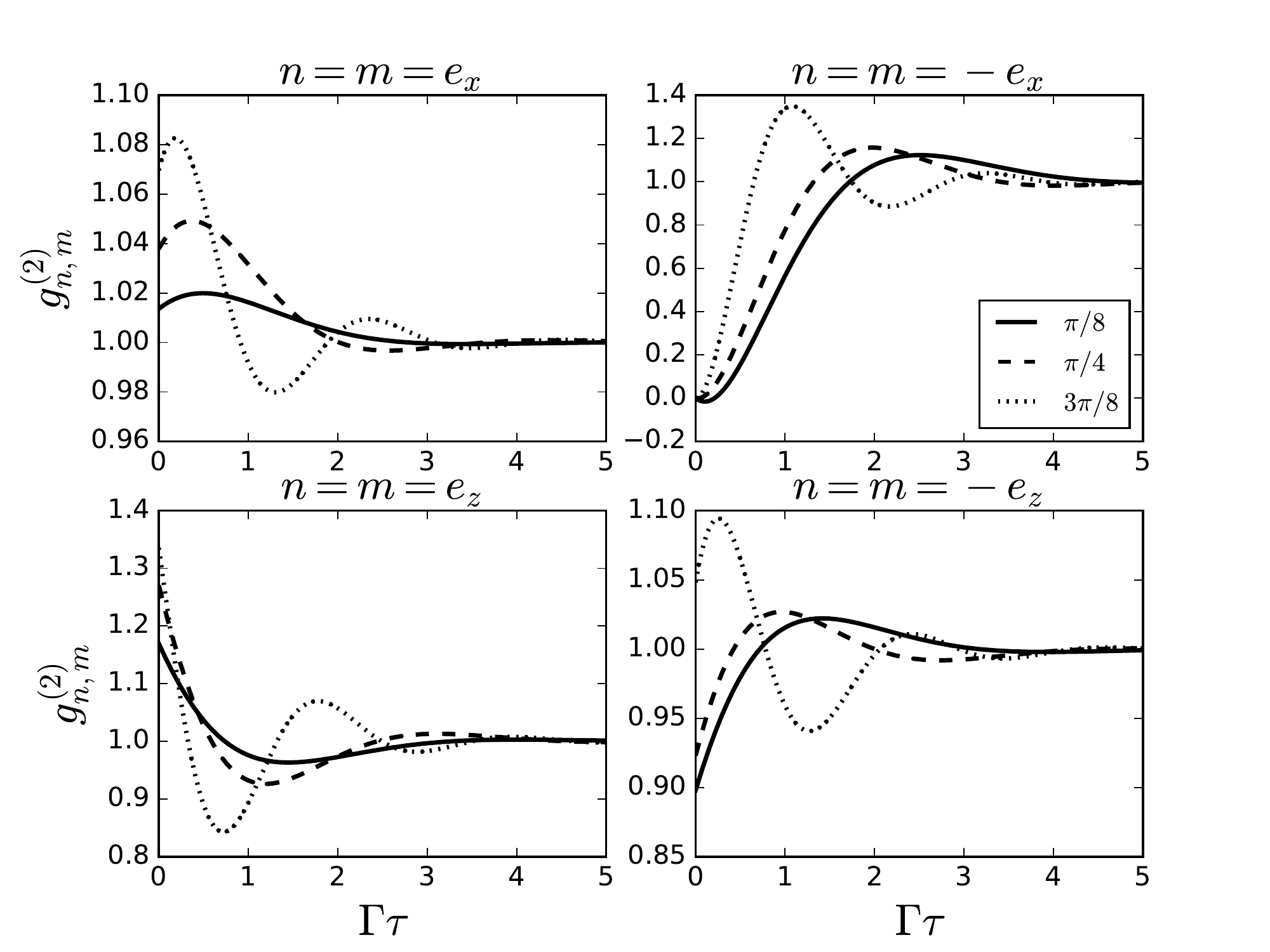}
\caption{Second order correlation function $g^{(2)}_{\boldsymbol{n},\boldsymbol{m}}  (\tau) $ for various values of $\psi$ (in the legend). The unit vectors $\boldsymbol{n}$ and $\boldsymbol{m}$ define  polarizations of the detectors used in coincidence measurements, and are aligned parallel to each other. The input signal is linearly polarized, $\boldsymbol{n}_{cl} = \boldsymbol{e}_x$.}
\label{par}
\end{figure}

\begin{figure}[t]
\centering
\includegraphics[width=90mm]{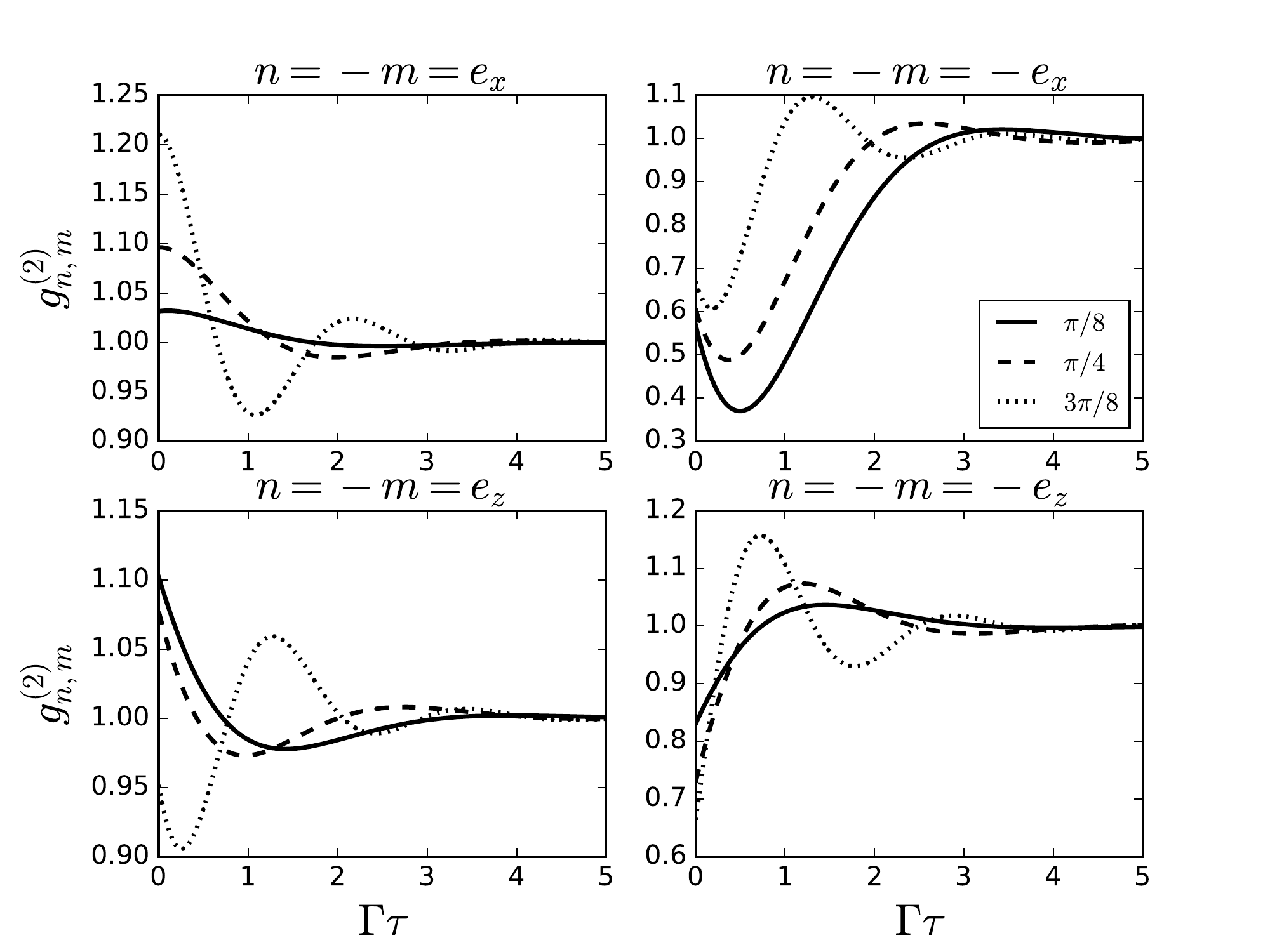}
\caption{Second order correlation function $g^{(2)}_{\boldsymbol{n},\boldsymbol{m}}  (\tau) $ for various values of $\psi$ (in the legend). The unit vectors $\boldsymbol{n}$ and $\boldsymbol{m}$ define  polarizations of the detectors used in coincidence measurements, and are aligned antiparallel to each other. The input signal is linearly polarized, $\boldsymbol{n}_{cl} = \boldsymbol{e}_x$.}
\label{antipar}
\end{figure}

In Figs. \ref{par} and \ref{antipar} we plot the function $g^{(2)}_{\boldsymbol{n},\boldsymbol{m}}  (\tau) $ for parallel and antiparallel alignments of the vectors $\boldsymbol{n}$ and $\boldsymbol{m}$, defining detectors' polarizations. We see that emitted photons, either with  linear or with circular polarizations, demonstrate diverse statistical features ranging from bunching to antibunching depending on the model parameters.

\section{Conclusion}
We derived the photonic analogue of the Kondo model in the far-detuned regime  of a three-level emitter coupled to waveguide modes with a linear dispersion and transitions between levels obeying angular momentum selection rules. The derived effective Hamiltonian coincides -- except for statistics of particles -- with the antiferromagnetic Kondo Hamiltonian arising in condensed matter models. 

Using the derived effective model we studied dynamics of the local system as well as various correlation functions  of scattered light assuming the initially coherent state. It turned out that in the photonic Kondo model all inelastic properties are tightly bound to a polarization of the initial state. In addition, they also quantify the degree of entanglement between the local system and the outgoing radiation. We proposed a way of experimentally establishing the model parameters performing various polarization resolved measurements. We studied the statistical properties of the outgoing radiation and observed that they are sensitive to both model parameters and the initial polarization. Moreover, the second order coherence shows oscillatory behavior and can possibly be used to engineer strongly correlated states of light.

We also observed that the quantum regression theorem holds in our model for the second order coherence connecting its dynamics to that of the local system. This is a consequence of the bilinear (in fields) light-matter coupling in the effective model. This feature is in contrast to standard applications of the quantum regression theorem to models with a linear coupling, where it relates local dynamics to the first order coherence.

\begin{acknowledgements}
The authors thank H. Schoeller and D. Schuricht for useful discussions.
The work of V. G. is part of the Delta-ITP consortium, a program of the Netherlands Organization for Scientific Research (NWO) funded by the Dutch Ministry of Education, Culture and Science (OCW). 
\end{acknowledgements}

\appendix
\section{Derivation of the effective Hamiltonian \eqref{dr8} by the Schrieffer-Wolff transformation}
\label{App:AppendixA}

The Schrieffer-Wolff transformation maps the three level Hamiltonian (\ref{dr1}), (\ref{dr2})  onto the effective two-level Hamiltonian
\begin{align}
H^{(3)} & \mapsto H^{(2)}  = P e^S H^{(3)} e^{-S} P \label{sw3}
 \\
&\approx P \lbrace H^{(3)} + \lbrack S , H^{(3)} \rbrack + \frac{1}{2} \lbrack S , \lbrack S , H^{(3)} \rbrack \rbrack \rbrace P ,\nonumber
\end{align}
where the generator $S$ is chosen in the form \eqref{S},
and $P\equiv\ket{+} \bra{+}+ \ket{-} \bra{-} $ is a projector onto the  subspace spanned by the states $ \ket{+}$ and $\ket{-}$.

Neglecting $\Delta$ against $\omega_3$ we find
\beq
\lbrack S , H^{(3)} \rbrack &=& \sum_{\sigma=\pm} \sigma  \int d \omega \Lambda_{\omega \sigma}  (\omega - \omega_3) \nonumber\\
& & \times (a_{\omega \sigma } \ket{3} \bra{-\sigma} + a^{\dagger}_{\omega \sigma} \ket{-\sigma} \bra{3})
\notag \\
&-& \sum_{\sigma, \sigma'=\pm}  \sigma \sigma' \int d \omega d \omega' (g_{\sigma} \Lambda_{\omega' \sigma'} +g_{\sigma'}  \Lambda_{\omega \sigma})\notag \\
& & \times  a^{\dagger}_{\omega' \sigma'} a_{\omega \sigma} \ket{-\sigma'} \bra{-\sigma} 
\label{sw7} \notag \\
&+ & \sum_{\sigma=\pm} \int d \omega d\omega' g_{\sigma} ( \Lambda_{\omega' \sigma} +  \Lambda_{\omega \sigma} ) \notag \\
& & \times a_{\omega \sigma} a^{\dagger}_{\omega' \sigma } \ket{3} \bra{3}
\label{sw8} 
\eeq
and
\begin{align}
& P \lbrack S , \lbrack S , H^{(3)} \rbrack \rbrack P  \notag \\
& \qquad = - 2 \sum_{\sigma, \sigma'} \sigma \sigma' \int d\omega  d\omega' \Lambda_{\omega' \sigma'} \Lambda_{\omega \sigma} \left(\frac{\omega' + \omega}{2} - \omega_3 \right) \notag \\
& \qquad  \qquad \times a^{\dagger}_{\omega' \sigma'} a_{\omega \sigma} \ket{-\sigma'} \bra{-\sigma}  .
\label{sw12} 
\end{align}
Choosing $\Lambda_{\omega \sigma} = \frac{g_{\sigma}}{\omega_3 - \omega}$ to suppress in \eqref{sw3} the terms linear in $a_{\omega \sigma}$ and $a_{\omega \sigma}^{\dagger}$, we arrive at the effective two-level Hamiltonian $H^{(2)}$ 
\begin{align}
H^{(2)} &= \Delta S_{z} + \sum_{\sigma} \int d\omega  \, \omega a^{\dagger}_{\omega \sigma} a_{\omega \sigma} \label{sw17}  \\
&- \frac{1}{2} \sum_{\sigma, \sigma' }  \sigma' \sigma \int d \omega \, d \omega' \, J_{\omega' \sigma', \omega \sigma} a^{\dagger}_{\omega' \sigma'} a_{\omega \sigma} \ket{-\sigma'} \bra{-\sigma},
\notag
\end{align}
where $J_{\omega'\sigma', \omega \sigma} = g_{\sigma'} g_{\sigma} (\frac{1}{\omega_3 - \omega'} + \frac{1}{\omega_3 -\omega}) $. This Hamiltonian can easily be transformed to the form (\ref{dr8}) under additional assumptions $g_+ = g_-  \equiv g$ and $J_{\omega' \sigma', \omega \sigma}= J_{\omega' \omega}  \approx J_{\omega_3/2 , \omega_3/2} = \frac{4 g^2}{\omega_3} \equiv J$. 

\section{Solving equations of motion}
\label{App:EqM}

Formally integrating \eqref{dr12} we obtain
\begin{align}
a_{\omega \sigma}(t) &= e^{-i \omega t} [ a_{\omega \sigma}(0) \nonumber\\
&- i J \int^t_0 d t' \, e^{i \omega t'} \int d \omega' \, M_{\sigma \sigma'}(t') a_{\omega' \sigma' }(t') ] .
\label{ps3}
\end{align}
Transforming \eqref{ps3} to the coordinate representation we find
\begin{align}
a_{\sigma}(x,t) &= a_{\sigma}(x-t,0)\label{ps4} \\
&- 2 \pi i J \Theta(x) \Theta(t-x) \sum_{\sigma'} M_{\sigma \sigma^{\prime}}(t-x) a_{\sigma^{\prime}}(0,t-x). 
 \notag
\end{align}
Setting $x=0$ and assuming $t > 0$ we obtain
\begin{equation}
a_{\sigma}(0,t) =\sum_{\sigma'} [1 + i \pi J M (t)]_{\sigma \sigma'}^{-1} a_{\sigma'}(-t,0).
\label{ps5}
\end{equation} 
Noticing that
\begin{align}
\sum_{\sigma'} M_{\sigma \sigma^{\prime}}(t) M_{\sigma^{\prime} \sigma^{\prime \prime}}(t) = - M_{\sigma \sigma^{\prime \prime}}(t),
\label{ps6}
\end{align}
we can express \eqref{ps5} as
\begin{align}
a_{\sigma}(0,t) = \sum_{\sigma'} \left[ 1 - \frac{i \pi J}{1 - i \pi J} M (t) \right]_{\sigma \sigma'} a_{\sigma^{\prime}}(-t,0).
\label{ps7}
\end{align}
Substituting this result into \eqref{ps4}  and considering $t > x > 0$ we obtain
\begin{align}
a_{\sigma}(x,t) = \sum_{\sigma'} [1+ (1 - e^{i \phi}) M (t-x) ]_{\sigma \sigma'} a_{\sigma'}(x-t,0),
\label{ps8}
\end{align}
where $e^{i \phi} =\frac{1+ i \pi J}{1- i \pi J}$.

The equation of motion for the spin operators $S^i(t)$ reads
\begin{align}
\frac{d}{dt} S^i (t) &= 2 \pi J \epsilon_{i j k} \sum_{\sigma, \sigma'} a^{\dagger}_{\sigma}(0,t) \frac{\sigma^j_{\sigma \sigma'}}{2} S^k (t) a_{\sigma^{\prime}}(0,t) \nonumber\\
&+ \Delta \epsilon_{i z k}  S^k (t).
\label{ps9}
\end{align}
Substituting \eqref{ps7} into \eqref{ps9} we obtain Eq.~\eqref{dr15}.

\section{Evaluation of polarization unresolved spectral function} 
\label{App:AppendixC}

The polarization unresolved correlation function is defined in \eqref{dr32}. In the stationary limit $t \to \infty$, its Fourier transform
\begin{align}
\tilde{C}^0 (\omega) =2 \, \text{Re} \int_0^{\infty} d\tau e^{-i \omega \tau} C^0 (\tau)
\label{FTun}
\end{align}
defines  the spectral function (or the power spectrum).

Defining the spin correlators $C^{ij}(\tau) = \braket{S^i(t+\tau) S^j(t)}$, we establish  equations for them  with help of \eqref{dr15} and \eqref{ps7}
\begin{align}
\frac{d}{d\tau} C^{ij}(\tau) &= \Omega \epsilon_{ilk} n^l_{h} C^{kj}(\tau) - \Gamma \left[ C^{ij}(\tau) - \frac{n^i_{cl} \braket{S^j }_{st}}{2}  \right]
\nonumber\\
&+ 2 \pi J \cos^2 \frac{\phi}{2} \sum_{\sigma, \sigma'} \epsilon_{ilk}  \langle a^{\dagger}_{\sigma}(-t_{\tau},0)  \frac{\sigma^l_{\sigma \sigma^{\prime}}}{2} S^k(t_{\tau}) \notag \\
& \times [ a_{\sigma'}(-t_{\tau},0) , S^j(t)] \rangle
\notag \\
&+ \frac{\pi J \sin \phi}{2} \sum_{\sigma, \sigma'} \langle a^{\dagger}_{\sigma}(- t_{\tau},0) \left[\frac{\sigma^i_{\sigma \sigma^{\prime}}}{2} - \delta_{\sigma \sigma'} S^i(t_{\tau}) \right] \notag \\
& \times [a_{\sigma^{\prime}}(-t_{\tau},0) , S^j(t)]\rangle,
\label{ps34} 
\end{align}
where $t_{\tau}= t + \tau$.

Let us consider the commutators appearing in the above relation
\begin{align}
&\lbrack a_{\sigma}(-t_{\tau},0) , S^i(t) \rbrack = \int \frac{d \omega}{\sqrt{2 \pi}} e^{- i \omega t_{\tau}} \lbrack  a_{\omega \sigma}(0) , S^i (t) \rbrack \notag \\
&= i J\int_{0}^{t}dt'\int \frac{d\omega d\omega'}{\sqrt{2 \pi}} e^{i \omega (t'-t_{\tau})}  
\lbrack M_{\sigma \sigma^{\prime}}(t^{\prime}) a_{\omega' \sigma'}(t^{\prime}), S^i(t)\rbrack 
\notag \\
&= \sqrt{2 \pi} i J\int_{0}^{t}dt'  \delta (t'-t_{\tau})  \int d\omega'  
\lbrack M_{\sigma \sigma^{\prime}}(t^{\prime}) a_{\omega' \sigma'}(t^{\prime}), S^i(t)\rbrack 
\notag \\
&= 2 \pi i  J \Theta(-\tau)  \lbrack M_{\sigma \sigma^{\prime}}(t_{\tau}) a_{\sigma^{\prime}}(0,t_{\tau}), S^i(t)\rbrack ,
\label{ps37}
\end{align}
where we have used (\ref{ps3}) and the commutation of the Heisenberg operators at equal time arguments, $\lbrack a_{\omega \sigma}(t) , S^i (t)\rbrack = 0$, in the second step. 

For $\tau > 0$ it follows
\begin{align}
\lbrack a_{\sigma}(-t_{\tau},0) , S^i (t)\rbrack = 0
\label{ps38}
\end{align}
and hence
\begin{align}
\frac{d}{d\tau} C^{ij}(\tau) = \Omega \epsilon_{ilk} n^l_{h} C^{kj}(\tau) - \Gamma \left[C^{ij}(\tau) - \frac{n^i_{cl} \braket{Sj}_{st}}{2} \right].
\label{ps43}
\end{align}

Defining
\begin{align}
A(\tau) &= \braket{\boldsymbol{S}(t + \tau) \cdot \boldsymbol{S}(t)} = C^{ii}(\tau),\\
B_{cl}(\tau) &= \boldsymbol{n}_{cl} \cdot \braket{\boldsymbol{S}(t + \tau) \times \boldsymbol{S}(t)} \notag \\
&= \epsilon_{ijk} n^k_{cl} C^{ij}(\tau),\\
B_h(\tau) &= \boldsymbol{n}_h \cdot \braket{\boldsymbol{S}(t + \tau) \times \boldsymbol{S}(t)} = \epsilon_{ijk} n^k_{h}  C^{ij}(\tau),\\
C_{cl,h}(\tau) &= \braket{(\boldsymbol{n}_{cl} \cdot \boldsymbol{S}(t + \tau)) ( \boldsymbol{n}_{h} \cdot \boldsymbol{S}(t))} \notag \\
&= n^i_{cl} n^j_{h} C^{ij}(\tau),
\end{align}
\begin{align}
C_{h,h}(\tau) &=\braket{(\boldsymbol{n}_h \cdot \boldsymbol{S}(t + \tau))(\boldsymbol{n}_h \cdot \boldsymbol{S}(t))} \notag \\
&= n^i_h n^j_h C^{ij}(\tau), \\
D(\tau) &=\braket{(\boldsymbol{n}_{cl} \cdot (\boldsymbol{n}_h \times \boldsymbol{S}(t + \tau)))(\boldsymbol{n}_h \cdot \boldsymbol{S}(t))}  \notag \\
&= \epsilon_{ikl} n^k_{cl} n^l_h n^j_h C^{ij}(\tau), 
\end{align}
we derive the following equations
\begin{align}
& \frac{d}{d\tau} A(\tau) = \Omega B_h(\tau) -  \Gamma \left[ A(\tau) - \frac{1}{4} \frac{1 + \lambda^2 \cos^2 \psi}{1 + \lambda^2}\right],
\label{ps50} \\
& \frac{d}{d\tau} B_{cl}(\tau) = \Omega [C_{cl,h}(\tau) - \cos \psi \, A(\tau)] -  \Gamma B_{cl}(\tau),
\label{ps51} \\
&\frac{d}{d\tau} B_h(\tau) = \Omega [C_{h,h}(\tau) - A(\tau)] \notag \\
&-  \Gamma \left[ B_h(\tau) - \frac{1}{4} \frac{\lambda \sin^2 \psi}{1 + \lambda^2} \right],
\label{ps52} \\
& \frac{d}{d\tau} C_{cl,h}(\tau) = \Omega D(\tau) -  \Gamma \left[C_{cl,h}(\tau) - \frac{\cos \psi}{4} \right],
\label{ps53} \\
& \frac{d}{d\tau} C_{h,h}(\tau) = - \Gamma \left[ C_{h,h}(\tau) - \frac{\cos^2 \psi}{4} \right],
\label{ps54} \\
& \frac{d}{d\tau} D(\tau) = \Omega [\cos \psi \, C_{h,h}(\tau) - C_{cl,h}(\tau)] -  \Gamma D(\tau),
\label{ps55}
\end{align}
with initial conditions specified by 
\begin{align}
A(0) &= \frac{3}{4},
\label{ps56} \\
B_{cl}(0) &= \frac{i}{2} \frac{1 + \lambda^2 \cos^2 \psi}{1 + \lambda^2},
\label{ps57} \\
B_h (0) &=  \frac{i}{2} \cos \psi,
\label{ps58} \\
C_{cl,h} (0) &=  \frac{\cos \psi}{4} - \frac{i}{4} \frac{\lambda \sin ^2 \psi}{1 + \lambda^2}, 
\label{ps59} \\
C_{h,h} (0) &= \frac{1}{4},
\label{ps60} \\
D (0) &= - \frac{i}{4} \frac{\sin^2 \psi}{1 + \lambda^2}.
\label{ps61} 
\end{align}

The solution of $C_{h,h}(\tau)$ can easily be found 
\begin{align}
C_{h,h}(\tau) &= \frac{1}{4} (\cos^2 \psi + \sin^2 \psi \, e^{- \Gamma \tau}).
\label{ps62}
\end{align}
Using this result we find
\begin{align}
& C_{cl,h}(\tau) = \frac{\cos \psi}{4} \left(\frac{1 + \lambda^2 \cos^2 \psi}{1 + \lambda^2} + \sin^2 \psi e^{- \Gamma \tau} \right) \notag  \\
&- \frac{i \sin^2 \psi}{4} \left( \cos^2 \frac{\psi}{2} \frac{e^{- (\Gamma - i \Omega) \tau}}{ \lambda + i}  + \sin^2 \frac{\psi}{2} \frac{e^{-(\Gamma + i \Omega) \tau}}{\lambda - i} \right), \\
& D(\tau) = \frac{\sin^2 \psi}{4} \left( \cos^2 \frac{\psi}{2} \frac{e^{-(\Gamma - i \Omega)\tau}}{\lambda + i} -  \sin^2 \frac{\psi}{2} \frac{e^{-(\Gamma + i \Omega) \tau}}{\lambda - i} \right. \notag \\
& \left.  \qquad \qquad- \frac{\lambda \cos \psi}{1 + \lambda^2} \right),
\label{ps65}
\end{align}
\begin{align}
A(\tau) &= \frac{1}{4} \frac{1 + \lambda^2 \cos^2 \psi}{1 + \lambda^2} + \frac{\sin^2 \psi}{4} e^{- \Gamma \tau}
\notag \\
&+ \frac{1}{2} \left(\cos^2 \frac{\psi}{2} - \frac{1}{4} \frac{\sin^2 \psi}{1 + \lambda^2} \right) e^{-(\Gamma - i \Omega) \tau} \notag \\
&+ \frac{1}{2} \left(\sin^2 \frac{\psi}{2} - \frac{1}{4} \frac{\sin^2 \psi}{1 + \lambda^2} \right) e^{-(\Gamma + i \Omega) \tau},
\label{ps67} 
\end{align}
\begin{align}
B_h(\tau) &= \frac{i}{2}  \left(\cos^2 \frac{\psi}{2} - \frac{1}{4} \frac{\sin^2 \psi}{1 + \lambda^2}\right) e^{-(\Gamma - i \Omega) \tau} \notag \\
&  -  \frac{i}{2} \left( \sin^2 \frac{\psi}{2} - \frac{1}{4} \frac{\sin^2 \psi}{1 + \lambda^2} \right) e^{-(\Gamma + i \Omega) \tau} .
\label{ps68}
\end{align}
This information allows us to find
\begin{align}
B_{cl}(\tau) &= \frac{i}{4} \frac{\sin^2 \psi}{1 + \lambda^2} (1 - i \lambda \cos \psi) e^{- \Gamma \tau}
\notag \\
&+ i \left[ \cos^2 \frac{\psi}{2} \left(\frac{\sin^2 \psi}{4} \frac{1}{1 - i \lambda} + \frac{\cos \psi}{2} \right) \right. \notag \\
&\left. - \frac{1}{8} \frac{\cos \psi \sin^2 \psi}{1 + \lambda^2} \right] e^{- (\Gamma - i \Omega)\tau}
\notag \\
&+ i \left[ \sin^2 \frac{\psi}{2} \left(\frac{\sin^2 \psi}{4} \frac{1}{1 + i \lambda} - \frac{\cos \psi}{2} \right) \right. \notag \\
& \left. + \frac{1}{8} \frac{\cos \psi \sin^2 \psi}{1 + \lambda^2} \right] e^{- (\Gamma + i \Omega)\tau}
\label{ps71}
\end{align}
and 
\begin{align}
A(\tau) + i B_{cl}(\tau) &= \frac{1}{4} \frac{1 + \lambda^2 \cos^2 \psi}{1 + \lambda^2} \label{ps74}\\
&+ \frac{1}{4} \frac{\lambda \sin^2 \psi}{1 + \lambda^2}(\lambda + i \cos \psi) e^{- \Gamma \tau} 
\notag \\
&+\frac{1}{4} \frac{\lambda \sin^2 \psi}{1 + \lambda^2} \left(\lambda - i \cos^2 \frac{\psi}{2} \right) e^{-(\Gamma - i \Omega) \tau}
\notag \\
&+ \frac{1}{4} \frac{\lambda \sin^2 \psi}{1 + \lambda^2} \left(\lambda + i \sin^2 \frac{\psi}{2} \right) e^{-(\Gamma + i \Omega) \tau}. \notag
\end{align}

Combining \eqref{dr32}, \eqref{FTun}, and \eqref{ps74} we obtain the polarization unresolved power spectrum \eqref{PS}.

\section{Evaluation of polarization resolved spectral function }
\label{App:corr-res}

To evaluate the polarization resolved correlation function \eqref{dr38} we additionally introduce
\beq
\boldsymbol{A}_L(\tau) &\equiv &\braket{(\boldsymbol{n}_{cl} \cdot \boldsymbol{S}(t + \tau)) \boldsymbol{S}(t)} ,
\label{pr5} \\
\boldsymbol{A}_R(\tau) &\equiv &\braket{\boldsymbol{S}(t + \tau) (\boldsymbol{n}_{cl} \cdot \boldsymbol{S}(t))},
\label{pr6} \\
\boldsymbol{B}_g(\tau) &\equiv &\braket{\boldsymbol{S}(t + \tau) \times \boldsymbol{S}(t)}.
\label{pr7} 
\eeq
Assuming $\boldsymbol{n}_{cl} \nparallel \boldsymbol{n}_h$ we expand \eqref{pr5}-\eqref{pr7} in the (nonorthogonal) basis $\{\boldsymbol{n}_{cl}, \boldsymbol{n}_h, \boldsymbol{n}_{cl} \times \boldsymbol{n}_h \}$ :
\begin{align}
\boldsymbol{A}_L(\tau) &= \frac{C_{cl,cl}(\tau) - \cos \psi \, C_{cl,h}(\tau)}{\sin^2 \psi} \boldsymbol{n}_{cl} \notag \\
&+ \frac{C_{cl,h}(\tau) - \cos \psi \, C_{cl,cl}(\tau)}{\sin^2 \psi} \boldsymbol{n}_h 
\notag \\
&+ \frac{E_R(\tau)}{\sin^2 \psi} \boldsymbol{n}_{cl} \times \boldsymbol{n}_h,
\label{pr9} 
\end{align}
\begin{align}
\boldsymbol{A}_R(\tau) &= \frac{C_{cl,cl}(\tau) - \cos \psi \, C_{h,cl}(\tau)}{\sin^2 \psi} \boldsymbol{n}_{cl} \notag \\
&+ \frac{C_{h,cl}(\tau) - \cos \psi \, C_{cl,cl}(\tau)}{\sin^2 \psi} \boldsymbol{n}_h
\notag \\ 
&+ \frac{E_L(\tau)}{\sin^2 \psi} \boldsymbol{n}_{cl} \times \boldsymbol{n}_h,
\label{pr11} 
\end{align}
\begin{align}
\boldsymbol{B}_g(\tau) &= \frac{B_{cl}(\tau) - \cos \psi \, B_h(\tau)}{\sin^2 \psi} \boldsymbol{n}_{cl} \notag \\
&+ \frac{B_h(\tau) - \cos \psi \, B_{cl}(\tau)}{\sin^2 \psi} \boldsymbol{n}_h 
\notag \\
&+ \frac{C_{cl,h}(\tau) - C_{h,cl}(\tau)}{\sin^2 \psi} \boldsymbol{n}_{cl} \times \boldsymbol{n}_h,
\label{pr13}
\end{align}
where
\begin{align}
C_{cl,cl}(\tau) &= \braket{(\boldsymbol{n}_{cl} \cdot \boldsymbol{S}(t + \tau)) (\boldsymbol{n}_{cl} \cdot \boldsymbol{S}(t))} \nonumber\\
&= n^i_{cl} n^j_{cl} C_{ij}(\tau),
\label{pr14} \\
C_{h,cl}(\tau) &= \braket{(\boldsymbol{n}_{h} \cdot \boldsymbol{S}(t + \tau)) (\boldsymbol{n}_{cl} \cdot \boldsymbol{S}(t))} \nonumber\\
&= n^i_h n^j_{cl} C_{ij}(\tau),
\label{pr15} \\
E_R(\tau) &= \braket{(\boldsymbol{n}_{cl} \cdot \boldsymbol{S}(t + \tau)) \boldsymbol{n}_{cl} \cdot (\boldsymbol{n}_h \times \boldsymbol{S}(t))} \nonumber\\
&= \epsilon_{jkl} n^k_{cl} n^i_{cl} n^l_h C_{ij}(\tau),
\label{pr16} \\
E_L(\tau) &= \braket{\boldsymbol{n}_{cl} \cdot (\boldsymbol{n}_h \times \boldsymbol{S}(t + \tau))(\boldsymbol{n}_{cl} \cdot \boldsymbol{S}(t)) } \nonumber\\
&= \epsilon_{ikl} n^k_{cl} n^j_{cl} n^l_h C_{ij}(\tau),
\label{pr17} \\
F(\tau) &= \braket{((\boldsymbol{n}_{cl} \times \boldsymbol{n}_h) \cdot \boldsymbol{S}(t + \tau)) ((\boldsymbol{n}_{cl} \times \boldsymbol{n}_h) \cdot \boldsymbol{S}(t))} \nonumber\\
&= \epsilon_{inm} \epsilon_{jkl} n^k_{cl} n^n_{cl} n^l_{h} n^m_{h} C_{ij}(\tau),
\label{pr18} \\
\bar{F}(\tau) &= \braket{(\boldsymbol{n}_h \cdot \boldsymbol{S}(t + \tau)) ((\boldsymbol{n}_{cl} \times \boldsymbol{n}_h) \cdot \boldsymbol{S}(t))}\nonumber\\
& = \epsilon_{jkl} n^k_{cl} n^l_h n^i_h C_{ij}(\tau).
\label{pr19}
\end{align}

From \eqref{ps43} we derive  the following equations at $\tau > 0$
\begin{align}
& \frac{d}{d\tau} C_{cl,cl}(\tau) =  \Omega E_L(\tau) - \Gamma \left[ C_{cl,cl}(\tau) - \frac{1}{4} \frac{1 + \lambda^2 \cos^2 \psi}{1 + \lambda^2}\right],
\label{pr20} \\
& \frac{d}{d\tau} C_{h,cl}(\tau) = - \Gamma \left[ C_{h,cl}(\tau) - \frac{\cos \psi}{4} \frac{1 + \lambda^2 \cos^2 \psi}{1 + \lambda^2}\right],
\label{pr21} \\
& \frac{d}{d\tau} E_R(\tau) = \Omega F(\tau) - \Gamma \left[ E_R(\tau) + \frac{1}{4} \frac{\lambda \sin^2 \psi}{1 + \lambda^2}\right],
\label{pr22} \\
& \frac{d}{d\tau} F (\tau) = \Omega [\cos \psi \, \bar{F}(\tau) - E_R(\tau)] - \Gamma F(\tau),
\label{pr23} \\
& \frac{d}{d\tau}  \bar{F} (\tau) = - \Gamma \left[\bar{F}(\tau) + \frac{\cos \psi }{4} \frac{\lambda \sin^2 \psi}{1 + \lambda^2} \right],
\label{pr24} \\
& \frac{d}{d\tau} E_L (\tau) = \Omega \left[\cos \psi \, C_{h,cl}(\tau) - C_{cl,cl}(\tau)\right] - \Gamma E_L(\tau),
\label{pr25}
\end{align}
with the  initial conditions
\begin{align}
C_{cl,cl}(0) &=\frac{1}{4},
\label{pr26} \\
C_{h,cl}(0) &= \frac{\cos \psi}{4} + \frac{i}{4} \frac{\lambda \sin^2 \psi}{1 + \lambda^2},
\label{pr27} \\
E_R(0) &= -i \frac{\cos \psi}{4} \frac{\lambda^2 \sin^2 \psi}{1 + \lambda^2},
\label{pr28} \\
F(0) &= \frac{\sin^2 \psi}{4},
\label{pr29} \\
\bar{F}(0) &= \frac{i}{4} \frac{\sin^2 \psi}{1 + \lambda^2},
\label{pr30} \\
E_L(0) &= i \frac{\cos \psi}{4} \frac{\lambda^2 \sin^2 \psi}{1 + \lambda^2}.
\label{pr31} 
\end{align}

We can immediately solve (\ref{pr21}) and find
\begin{align}
C_{h,cl}(\tau) & = \frac{\cos \psi}{4} \frac{1 + \lambda^2 \cos^2 \psi}{1 + \lambda^2} \nonumber\\
&+ \frac{1}{4} \frac{\lambda \sin^2 \psi}{1 + \lambda^2} (\lambda \cos \psi + i) e^{- \Gamma \tau}.
\label{pr32}
\end{align}
This result allows us to solve \eqref{pr20} and \eqref{pr25} in the next step. Thus we obtain
\begin{align}
C_{cl,cl}(\tau) &= \frac{1}{4} \left(\frac{1 + \lambda^2 \cos^2 \psi}{1 + \lambda^2} \right)^2 \label{pr33}  \\
&+ \frac{\cos \psi}{4} \frac{\lambda \sin^2 \psi}{1 + \lambda^2} (\lambda \cos \psi + i) e^{- \Gamma \tau}
\notag \\
&+ \frac{1}{8} \frac{\lambda \sin^2 \psi}{1 + \lambda^2} \left[ \frac{\lambda \sin^2 \psi}{1 + \lambda^2} + \lambda (1 + \cos \psi)
\right. \notag \\
& \left. - i \left(\cos \psi + \frac{1 + \lambda^2 \cos^2 \psi}{1 + \lambda^2} \right) \right] e^{-(\Gamma - i \Omega) \tau}
\notag \\
&+ \frac{1}{8} \frac{\lambda \sin^2 \psi}{1 + \lambda^2} \left[ \frac{\lambda \sin^2 \psi}{1 + \lambda^2} + \lambda (1 - \cos \psi)
\right. \notag \\
& \left. - i \left(\cos \psi - \frac{1 + \lambda^2 \cos^2 \psi}{1 + \lambda^2} \right) \right] e^{-(\Gamma + i \Omega) \tau},\nonumber
\end{align}
\begin{align}
E_L(\tau) &= - \lambda \frac{\sin^2 \psi}{4} \frac{1 + \lambda^2 \cos^2 \psi}{(1 + \lambda^2)^2} \label{pr34}  \\
&+ \frac{i}{8} \frac{\lambda \sin^2 \psi}{1 + \lambda^2} \left[ \frac{\lambda \sin^2 \psi}{1 + \lambda^2} + \lambda (1 + \cos \psi)
\right. \notag \\
& \left. - i \left( \cos \psi + \frac{1 + \lambda^2 \cos^2 \psi}{1 + \lambda^2} \right) \right] e^{-(\Gamma - i \Omega) \tau}
\notag \\
&- \frac{i}{8} \frac{\lambda \sin^2 \psi}{1 + \lambda^2} \left[ \frac{\lambda \sin^2 \psi}{1 + \lambda^2} + \lambda (1 - \cos \psi)
\right. \notag \\
& \left. - i \left(\cos \psi - \frac{1 + \lambda^2 \cos^2 \psi}{1 + \lambda^2} \right) \right] e^{-(\Gamma + i \Omega) \tau}.
\notag
\end{align}

Similarly, from \eqref{pr24} we find
\begin{align}
\bar{F}(\tau) &= \frac{1}{4} \frac{\sin^2 \psi}{1 + \lambda^2} \left[ - \lambda \cos \psi + (i + \lambda \cos \psi) e^{- \Gamma \tau} \right],
\label{pr35}
\end{align}
which  enables us  to solve \eqref{pr22} and \eqref{pr23}. Thus we find
\begin{align}
F(\tau) &= \frac{1}{4} \frac{\lambda^2 \sin^4 \psi}{(1 + \lambda^2)^2} \label{pr36} \\
&+ \frac{\sin^2 \psi}{8} \left( 1 + \cos \psi + \frac{i \lambda}{1 - i \lambda} \frac{\sin^2 \psi}{1 + \lambda^2} \right) e^{-(\Gamma - i \Omega) \tau} \notag \\
&+ \frac{\sin^2 \psi}{8} \left( 1 - \cos \psi - \frac{i \lambda}{1 + i \lambda} \frac{\sin^2 \psi}{1 + \lambda^2}  \right) e^{-(\Gamma + i \Omega) \tau} , \notag
\end{align}
and
\begin{align}
E_R(\tau) &= - \frac{1}{4} \frac{\lambda \sin^2 \psi \, (1 + \lambda^2 \cos^2 \psi)}{(1 + \lambda^2)^2} \label{pr37} \\
&+ \frac{1}{4} \frac{\cos \psi \, \sin^2 \psi}{1 + \lambda^2} (i + \lambda \cos \psi) e^{-\Gamma \tau}\notag \\
&- \frac{i \sin^2 \psi}{8} \left( 1 + \cos \psi + \frac{i \lambda}{1 - i \lambda} \frac{\sin^2 \psi}{1 + \lambda^2} \right) e^{-(\Gamma - i \Omega)\tau} \notag \\
&+ \frac{i \sin^2 \psi}{8} \left( 1 - \cos \psi - \frac{i \lambda}{1 + i \lambda} \frac{\sin^2 \psi}{1 + \lambda^2} \right) e^{-(\Gamma + i \Omega) \tau}.   \notag
\end{align}

Collecting all contributions we obtain
\begin{align}
\boldsymbol{A}_R(\tau) &+ \boldsymbol{A}_L(\tau) - \boldsymbol{n}_{cl} A(\tau) - i \boldsymbol{B}_g(\tau) 
\notag \\
& = a \,   \boldsymbol{n}_{cl} + b \, \boldsymbol{n}_{h} + c \, \, \boldsymbol{n}_{cl} \times \boldsymbol{n}_h,
\label{pr40}
\end{align}
where the coefficients $a,b,c$ are given by
\begin{align}
a &= \frac{1}{4} \frac{(1 - \lambda^2)(1 + \lambda^2 \cos^2 \psi)}{(1 + \lambda^2)^2} - \frac{1}{4} \frac{\lambda^2 \sin^2 \psi}{1 + \lambda^2} e^{-\Gamma \tau} \label{pr41}  \\
&- \frac{i}{8} \frac{\lambda \sin^2 \psi}{1 + \lambda^2} \left[ \frac{1 + i \lambda}{1 - i \lambda} e^{- (\Gamma - i \Omega)\tau} - \frac{1 - i \lambda}{1 + i \lambda} e^{- (\Gamma + i \Omega)\tau} \right], \notag
\end{align}
\begin{align}
b &=\frac{ \lambda^2 \cos \psi}{2} \frac{1 + \lambda^2 \cos^2 \psi}{(1 + \lambda^2)^2} + \frac{1}{4} \frac{\lambda \sin^2 \psi}{1 + \lambda^2} (i + 2 \lambda \cos \psi) e^{- \Gamma \tau}
\notag \\
&+ \frac{1}{8} \frac{\lambda \sin^2 \psi}{1 + \lambda^2} \left[ \left(2 \lambda  - 2 \lambda \frac{\cos \psi}{1 - i \lambda} - i \right) e^{-(\Gamma - i \Omega) \tau} \right. \label{pr42} \\
& \left. - \left(2 \lambda  + 2 \lambda \frac{\cos \psi}{1 + i \lambda} + i \right) e^{-(\Gamma + i \Omega) \tau} \right],
\notag
\end{align}
\begin{align}
c &= - \frac{\lambda}{2} \frac{1 + \lambda^2 \cos^2 \psi}{(1 + \lambda^2)^2} - \frac{1}{4} \frac{\lambda \sin^2 \psi}{1 + \lambda^2} e^{- \Gamma \tau} \label{pr43}  \\
&+ \frac{1}{8} \frac{\lambda \sin^2 \psi}{1 + \lambda^2} \left[ \frac{1 + i \lambda}{1 - i \lambda} e^{-(\Gamma - i \Omega) \tau} + \frac{1 - i \lambda}{1 + i \lambda} e^{-(\Gamma + i \Omega) \tau} \right] .\notag
\end{align}

The Fourier transform of \eqref{pr40} leads us to \eqref{dr41}

\section{Computation of $g^{(2)}$ function}
\label{App:g2}

To obtain expression for the second-order correlation functions defined in \eqref{Gnm} and \eqref{Gn0}, we introduce the quantities
\begin{align}
\boldsymbol{K}_{\boldsymbol{m}}(\tau) &= \sum_{\tilde{\sigma}', \tilde{\sigma}} \sum_{\sigma', \sigma} (\boldsymbol{m} \cdot \frac{\boldsymbol{\sigma}_{\sigma' \sigma}}{2}) \frac{\alpha^*_{\tilde{\sigma}'} \alpha_{\tilde{\sigma}}}{L} \nonumber\\
&\times \langle [ \delta_{\tilde{\sigma}' \sigma' }+ (1-e^{- i \phi}) M_{\tilde{\sigma}' \sigma' } (t)]  \boldsymbol{S}(t + \tau)
\notag \\
&\times [ \delta_{\sigma \tilde{\sigma}}+ (1-e^{ i \phi}) M_{\sigma \tilde{\sigma}} (t)] \rangle,
\label{gt14} \\
\boldsymbol{K}_0(\tau) &= \sum_{\tilde{\sigma}', \tilde{\sigma}} \sum_{\sigma}  \frac{\alpha^*_{\tilde{\sigma}'} \alpha_{\tilde{\sigma}}}{L} \notag \\
&\times \langle [ \delta_{\tilde{\sigma}' \sigma} + (1-e^{ -i \phi}) M_{\tilde{\sigma}' \sigma} (t)] \boldsymbol{S}(t + \tau) 
\notag \\
&\times [ \delta_{ \sigma \tilde{\sigma}}+ (1-e^{ i \phi}) M_{\sigma \tilde{\sigma} } (t)] \rangle,
\label{gt15}
\end{align}
and using \eqref{dr15} we establish that they obey  the equations
\begin{align}
\frac{d}{d\tau} \boldsymbol{K}_{\boldsymbol{m}}(\tau) &= \Omega \boldsymbol{n}_h \times \boldsymbol{K}_{\boldsymbol{m}}(\tau) 
\label{gt16} \\
&- \Gamma [ \boldsymbol{K}_{\boldsymbol{m}}(\tau) - \frac{A_{\boldsymbol{m}}}{2} \boldsymbol{n}_{cl}  ],\nonumber
\\
\frac{d}{d\tau} \boldsymbol{K}_{0}(\tau) &= \Omega \boldsymbol{n}_h \times \boldsymbol{K}_{0}(\tau) - \Gamma [ \boldsymbol{K}_{0}(\tau) - \frac{A_0}{2} \boldsymbol{n}_{cl} ],
\label{gt17}
\end{align}
where
\begin{align}
A_{\boldsymbol{m}} &= \sum_{\tilde{\sigma}', \tilde{\sigma}} \sum_{\sigma', \sigma} (\boldsymbol{m} \cdot \frac{\boldsymbol{\sigma}_{\sigma' \sigma}}{2}) \frac{\alpha^*_{\tilde{\sigma}'} \alpha_{\tilde{\sigma}}}{L} \nonumber\\
&\times \langle [ \delta_{\tilde{\sigma}' \sigma' }+ (1-e^{- i \phi}) M_{\tilde{\sigma}' \sigma' } (t)]  
\notag \\
&\times [ \delta_{\sigma \tilde{\sigma}} + (1-e^{ i \phi}) M_{\sigma \tilde{\sigma}} (t)] \rangle = \boldsymbol{m} \cdot \boldsymbol{C}_{s} (0),
\label{gt14a} \\
A_0  &= \sum_{\tilde{\sigma}', \tilde{\sigma}} \sum_{\sigma}  \frac{\alpha^*_{\tilde{\sigma}'} \alpha_{\tilde{\sigma}}}{L} \notag \\
&\times \langle [ \delta_{\tilde{\sigma}' \sigma}+ (1-e^{ -i \phi} ) M_{\tilde{\sigma}' \sigma} (t)]  
\notag \\
&\times [ \delta_{ \sigma \tilde{\sigma}} + (1-e^{ i \phi}) M_{\sigma \tilde{\sigma} } (t)] \rangle =f.
\label{gt15a}
\end{align}

After rescaling $\boldsymbol{k}_{\boldsymbol{m}}(\tau) =\frac{\boldsymbol{K}_{\boldsymbol{m}}(\tau)}{A_{\boldsymbol{m}} }$
and $\boldsymbol{k}_0(\tau) = \frac{\boldsymbol{K}_0(\tau)}{f}$, we obtain the equations quoted in Section \ref{g2sect}.


\begin{thebibliography}{99}


\bibitem{BH1}
A. D. Greentree, C. Tahan, J. H. Cole, and L. C. L. Hollenberg, Quantum phase transitions of light, \href{http://dx.doi.org/10.1038/nphys466}{Nature Phys. {\bf 2}, 856 (2006). }

\bibitem{BH2}
M. J. Hartmann, F. G. S. L. Brandao, and M. B. Plenio, Strongly interacting polaritons in coupled arrays of cavities,\href{http://dx.doi.org/10.1038/nphys462}{Nature Phys. {\bf 2}, 849 (2006).} 

\bibitem{LL1}
D. E. Chang, V. Gritsev, G. Morigi, V. Vuletic, M. D. Lukin, and E. A. Demler, Crystallization of strongly interacting photons in a nonlinear optical fibre,  \href{http://dx.doi.org/10.1038/nphys1074}{Nature Phys. {\bf 4}, 884 (2008).}


\bibitem{LL2}
I. Carusotto, D. Gerace, H. E. Tureci, S. De Liberato, C. Ciuti, and A. Imamoglu, Fermionized Photons in an Array of Driven Dissipative Nonlinear Cavities, \href{http://dx.doi.org/10.1103/PhysRevLett.103.033601}{Phys. Rev. Lett. {\bf 103}, 033601 (2009).}




\bibitem{BEC}
J. Klaers, J. Schmitt,	 F. Vewinger,	and  M. Weitz, BoseÐEinstein condensation of photons in an optical microcavity, \href{http://dx.doi.org/10.1038/nature09567}{Nature {\bf 468}, 545 (2010).}


\bibitem{top1}
Z. Wang, Y. Chong, J. D. Joannopoulos, and  M. Solja\'{c}i\'{c}, Observation of unidirectional backscattering-immune topological electromagnetic states, \href{http://dx.doi.org/10.1038/nature08293}{Nature {\bf 461}, 772 (2009).}

\bibitem{top2}
A. B. Khanikaev, S. H. Mousavi, W.-K. Tse, M. Kargarian,	 A. H. MacDonald, and G. Shvets, Photonic topological insulators, \href{http://dx.doi.org/10.1038/nmat3520}{Nature Mat. {\bf 12}, 233 (2013).}


\bibitem{Hafezi1}
M. Hafezi, E. A. Demler, M. D. Lukin, and J. M. Taylor, Robust optical delay lines with topological protection, Nature Phys. {\bf 7}, 907 (2011).



\bibitem{CiutiRMP}
I. Carusotto and C. Ciuti, Quantum fluids of light, \href{http://dx.doi.org/10.1103/RevModPhys.85.299}{Rev. Mod. Phys. {\bf 85}, 299 (2013).}







\bibitem{Kimble}
H. J. Kimble,  The quantum internet, \href{http://dx.doi.org/10.1038/nature07127}{Nature {\bf 453}, 1023 (2008).}


\bibitem{ChuLukin}
Y. Chu and M. D. Lukin, Quantum optics with nitrogen-vacancy centers in diamond, 	arXiv:1504.05990.


\bibitem{Neumann}
P. Neumann, {\it et al}., Multipartate entanglement
among single spins in diamond, \href{http://dx.doi.org/0.1126/science.1157233}{Science {\bf 320}, 1326 (2008).}



\bibitem{Ansmann}
M. Ansmann, H. Wang, R. C. Bialczak, M. Hofheinz, E. Lucero, M. Neeley, A. D. O'Connell, D. Sank, M. Weides, J. Wenner, A. N. Cleland, and J. M. Martinis, 
Violation of Bell's inequality in Josephson phase qubits, \href{http://dx.doi.org/10.1038/nature08363}{Nature {\bf 461}, 504 (2009).}



\bibitem{JW-room}
F. Dolde,	 I. Jakobi, B. Naydenov, N. Zhao, S. Pezzagna,	 C. Trautmann,	J. Meijer, P. Neumann, F. Jelezko,	 and J. Wrachtrup,
Room-temperature entanglement between single defect spins in diamond, \href{http://dx.doi.org/doi:10.1038/nphys2545}{Nature Physics {\bf 9}, 139 (2013).}



\bibitem{Bernien}
H. Bernien,	 B. Hensen,	 W. Pfaff,	 G. Koolstra,	 M. S. Blok,	 L. Robledo,	 T. H. Taminiau,	 M. Markham, D. J. Twitchen, L. Childress, and R. Hanson, Heralded entanglement between solid-state qubits separated by three meters, \href{http://dx.doi.org/10.1038/nature12016}{Nature {\bf 497},  86 (2013).}




\bibitem{Kondo}
J. Kondo, Resistance minimum in dilute magnetic alloys, \href{http://dx.doi.org/10.1143/PTP.32.37}{Prog. Theor. Phys. {\bf  32}, 37 (1964).}

\bibitem{Anderson}
P. W. Anderson, Localized magnetic states in metals, \href{http://dx.doi.org/10.1103/PhysRev.124.41}{Phys. Rev., {\bf 124}, 41 (1961).}

\bibitem{Cronenwett}
S. M. Cronenwett, T. H. Oosterkamp, and L. P. Kouwenhoven, A tunable Kondo effect in quantum dots, \href{http://dx.doi.org/10.1126/science.281.5376.540}{Science {\bf 281}, 540 (1998).}

\bibitem{Glazman}
L. I. Glazman and M. E. Raikh, Resonant Kondo transparency of a barrier with quasilocal impurity states, PisÕma Zh. Eksp. Teor. Fiz. {\bf 47}, 378 (1988).









\bibitem{Le1}
A. LeClair,   QED for a Fibrillar Medium of Two-Level Atoms, 
\href{http://dx.doi.org/10.1103/PhysRevA.56.782}{Phys. Rev. A {\bf 56}, 782 (1997).}

\bibitem{KoLe}
R. Konik and A. LeClair, The Scattering Theory of Oscillator Defects in an Optical Fiber, 
\href{http://dx.doi.org/10.1103/PhysRevB.58.1872}{Phys. Rev. B {\bf 58}, 1872 (1998).}

\bibitem{LLLS}
A. LeClair, F. Lesage, S. Lukyanov, and H. Saleur,  The Maxwell-Bloch Theory in Quantum Optics and the Kondo Model,
\href{http://dx.doi.org/10.1016/S0375-9601(97)00602-6}{Phys. Lett. A {\bf 235}, 203 (1997).}

\bibitem{Le2}
A. LeClair, Eigenstates of the Atom-Field Interaction and the Binding of Light in Photonic Crystals,
\href{http://dx.doi.org/10.1006/aphy.1998.5874}{Ann. Phys. {\bf 271}, 268 (1999).}

\bibitem{LeHur1}
K. Le Hur, Kondo resonance of a microwave photon, \href{http://dx.doi.org/10.1103/PhysRevB.85.140506}{
Phys. Rev. B {\bf 85}, 140506(R) (2012).}

\bibitem{LeHur2}
K. Le Hur, Quantum Phase Transitions in Spin-Boson Systems: Dissipation and Light Phenomena, in {\it Understanding Quantum Phase Transitions}, edited by L. D. Carr (Taylor and Francis, Boca Raton, 2010).


\bibitem{Tureci1}
H. E. T\"{u}reci, M. Hanl, M. Claassen, A. Weichselbaum, T. Hecht, B. Braunecker, A. Govorov, L. Glazman, A. Imamoglu, and J. von Delft, Many-Body Dynamics of Exciton Creation in a Quantum Dot by Optical Absorption: A Quantum Quench towards Kondo Correlations, \href{http://dx.doi.org/10.1103/PhysRevLett.106.107402}{Phys. Rev. Lett. {\bf 106}, 107402 (2011).}

\bibitem{Tureci2}
B. Sbierski, M. Hanl, A. Weichselbaum, H. E. T\"{u}reci, M. Goldstein, L. I. Glazman, J. von Delft, and A. Imamoglu, Proposed Rabi-Kondo Correlated State in a Laser-Driven Semiconductor Quantum Dot,
\href{http://dx.doi.org/10.1103/PhysRevLett.111.157402}{Phys. Rev. Lett. {\bf 111}, 157402 (2013).}



\bibitem{LukinEIT}
M. D. Lukin, Colloquium: Trapping and manipulating photon states in atomic ensembles, \href{http://dx.doi.org/10.1103/RevModPhys.75.457}{Rev. Mod. Phys. {\bf 75}, 457 (2003).}



\bibitem{FIM}
M. Fleischhauer, A. Imamoglu, and J. P. Marangos, Electromagnetically induced transparency: Optics in coherent
media, \href{http://dx.doi.org/10.1103/RevModPhys.77.633}{Rev. Mod. Phys. {\bf 77}, 633 (2005).}







\bibitem{AGMO}
G. Alzetta, A. Gozzini, L. Moi, and G. Orriols, An experimental method for the observation of r.f. transitions
and laser beat resonances in oriented na vapour, Il Nuovo Cimento B {\bf 36}, 5 (1976).



\bibitem{AO}
E. Arimondo and G. Orriols, Nonabsorbing atomic coherences by coherent two-photon transitions in a three level optical pumping, \href{http://dx.doi.org/10.1007/BF02746514}{Lettere Al Nuovo Cimento (1971 Ð1985) {\bf 17}, 333Ð338 (1976).}

\bibitem{GWS}
H. R. Gray, R. M. Whitley, and C. R. Stroud, Jr., Coherent trapping of atomic populations, \href{http://dx.doi.org/10.1364/OL.3.000218}{Optics Lett. {\bf 3}, 218 (1978). }

\bibitem{Orriols}
G. Orriols, Nonabsorption resonances by nonlinear coherent effects
in a three-level system, Il Nuovo Cimento {\bf 53} B, 1 (1979).

\bibitem{Arimondo}
E. Arimondo, Relaxation processes in coherent-population trapping, \href{http://dx.doi.org/10.1103/PhysRevA.54.2216}{Phys. Rev. A {\bf  54}, 2216 (1996).}

\bibitem{Aspect}
16. A. Aspect, E. Arimondo, R. Kaiser, N. Vansteenkiste, and C. Cohen-Tannoudji, Laser cooling below the one photon
recoil energy by velocity-selective coherent population trapping, \href{http://dx.doi.org/10.1103/PhysRevLett.61.826}{Phys. Rev. Lett. {\bf 61}, 826 (1988).}














\bibitem{stirap1}
Kuklinski, U. Gaubatz, Hioe, and K. Bergmann, Adiabatic population transfer in a three-level system driven by delayed laser pulses, \href{http://dx.doi.org/10.1103/PhysRevA.40.6741}{Phys. Rev. A {\bf 40}, 6741 (1989).}


\bibitem{stirap2}
U. Gaubatz, P. Rudecki, S. Schiemann, and K. Bergmann, Population transfer between molecular levels by stimulated Raman scattering with partially overlapping pulses: A new concept and experimental results, \href{http://dx.doi.org/10.1063/1.458514}{J. Chem. Phys. {\bf 92}, 5363 (1990).}


\bibitem{stirap-rev1}
K. Bergmann, H. Theuer, and B. W. Shore, Coherent population transfer among quantum states of atoms and molecules, \href{http://dx.doi.org/10.1103/RevModPhys.70.1003}{Rev. Mod. Phys. {\bf 70}, 1003 (1998).}

\bibitem{stirap-rev2}
N. V. Vitanov, M. Fleischhauer, B. W. Shore, and K. Bergmann, Coherent manipulation of atoms and molecules by sequential pulses, \href{http://dx.doi.org/10.1016/S1049-250X(01)80063-X}{Adv. in Atomic, Molecular and Optical Physics {\bf 46}, 55 (2001).}

\bibitem{stirap-rev4}
K. Bergmann, N. V. Vitanov, and B. W. Shore, Perspective: Stimulated Raman adiabatic passage: The status after 25 years, \href{http://dx.doi.org/10.1063/1.4916903}{J. Chem. Phys. {\bf 142}, 170901 (2015).}




\bibitem{AT}
S. H. Autler and C. H. Townes, Stark effect in rapidly varying fields, \href{http://dx.doi.org/10.1103/PhysRev.100.703}{Phys. Rev. {\bf 100}, 703 (1955).}

\bibitem{AT-CT}
C. Cohen-Tannoudji, The Autler-Townes effect revisited, in {\it Amazing Light}, (Springer, 1996).


\bibitem{EIT-vs-AT}
P. M. Anisimov, J. P. Dowling, and B. C. Sanders, Objectively discerning Autler-Townes splitting from electromagnetically induced transparency, \href{http://dx.doi.org/10.1103/PhysRevLett.107.163604}{
Phys. Rev. Lett. {\bf 107}, 163604 (2011).}



\bibitem{CT-R}
C. Cohen-Tannoudji and S. Reynaud, Dressed-atom description of resonance fluorescence and absorption spectra of a multi-level atom in an intense laser beam, \href{http://dx.doi.org/10.1088/0022-3700/10/3/005}{J. Phys. B: Atom. Molec. Phys., {\bf 10}, 345 (1977).}



\bibitem{WS}
R. M. Whitley and C. R. Stroud, Jr., Double optical resonance
\href{http://dx.doi.org/10.1103/PhysRevA.14.1498}{Phys. Rev. A {\bf 14}, 1498 (1976).}



\bibitem{Sobol}
B. Sobolewska, The fluorescence spectrum of a three-level atom in two-photon resonance with the laser field, \href{http://dx.doi.org/10.1016/0030-4018(77)90208-5}{Optics Comm. {\bf 20}, 378 (1977).}

\bibitem{Fu}
C.-R. Fu, Y.-M. Zhang, and C.-de Gong, Resonance fluorescence from a three-level system, \href{http://dx.doi.org/10.1103/PhysRevA.45.505}{Phys. Rev A. {\bf 45}, 505 (1992).}


\bibitem{Mavroyannis1}
C. Mavroyannis, Resonance fluorescence spectrum of a three-level atom, \href{http://dx.doi.org/10.1016/0030-4018(79)90142-1}{Optics Communications {\bf 29}, 80 (1979).}


\bibitem{Mavroyannis2}
C. Mavroyannis, Two-photon resonance fluorescence, \href{http://dx.doi.org/10.1016/0030-4018(78)90245-6}{
Optics Communications, {\bf 26}, 453 (1978).}



\bibitem{AB}
M. Alexanian and S. K. Bose, Two-photon resonance fluorescence, \href{http://dx.doi.org/10.1103/PhysRevA.74.063418}{Phys. Rev. A {\bf 74}, 063418 (2006).}


\bibitem{BPK}
S. K. Basu, T. Pramila, and D. Kanjilal, Dynamic Stark splitting in two-photon resonance fluorescence, \href{http://dx.doi.org/10.1016/0030-4018(83)90406-6}{Optics Communications
{\bf 45}, 43 (1983).}

\bibitem{exp-qd}
A. N. Vamivakas,Y. Zhao, C.-Y. Lu, and M. Atat\"{u}re, Spin-resolved quantum-dot resonance fluorescence,\href{http://dx.doi.org/10.1038/nphys1182}{Nature Physics {\bf 5}, 198 (2009). }



\bibitem{HH}
S. E. Harris and L. V. Hau, Nonlinear optics at low light levels, \href{http://dx.doi.org/10.1103/PhysRevLett.82.4611}{Phys. Rev. Lett. {\bf 82}, 4611 (1999).}

\bibitem{Kash}
M. M. Kash, V. A. Sautenkov, A. S. Zibrov, L. Hollberg, G. R.Welch, M. D. Lukin, Y. Rostovtsev, E. S. Fry, and
M. O. Scully, Ultraslow group velocity and enhanced nonlinear optical effects in a coherently driven hot atomic gas, \href{http://dx.doi.org/10.1103/PhysRevLett.82.5229}{Phys. Rev. Lett. {\bf 82}, 5229 (1999).}

\bibitem{WGX}
H. Wang, D. Goorskey, and M. Xiao, Enhanced Kerr nonlinearity via atomic coherence in a three-level atomic system,\href{http://dx.doi.org/10.1103/PhysRevLett.87.073601}{Phys. Rev. Lett. {\bf 87}, 073601 (2001).}


\bibitem{DSW}
W. Denk, J. Strickler, and W. Webb, Two-photon laser scanning fluorescence microscopy, \href{http://dx.doi.org/10.1126/science.2321027}{Science {\bf 248}, 76 (1990).}

\bibitem{2-photon}
P. T. C. So, C. Y. Dong, B. R. Masters, and K. M. Berland, Annu. Rev. Biomed. Eng. {\bf 2}, 399 (2000).






\bibitem{Budker}
D. Budker, W. Gawlik, D. F. Kimball, S. M. Rochester, V. V. Yashchuk, and A. Weis, Resonant nonlinear magneto-optical effects in atoms, \href{http://dx.doi.org/10.1103/RevModPhys.74.1153}{Rev. Mod. Phys. {\bf 74}, 1153 (2002).}





\bibitem{SchriefferWolff}
J. R. Schrieffer and P. A. Wolff, Relation between the Anderson and Kondo Hamiltonians, \href{http://dx.doi.org/10.1103/PhysRev.149.491}{Phys. Rev. {\bf 149}, 491 (1966).}



\bibitem{Jones}
R. Clark Jones, A new calculus for the treatment of optical systems: description and discussion of the calculus, \href{http://dx.doi.org/10.1364/JOSA.31.000488}{J. Opt. Soc. Am., {\bf 31}, 488 (1941).}

\bibitem{GardinerZoller}
P. Zoller and C. Gardiner, {\it Quantum Noise} (Springer-Verlag, 2004).




\end{thebibliography}
\end{document}